\documentclass[reprint,english,aps,prl,showpacs,floatfix,superscriptaddress]{revtex4-1}
\usepackage{inputenc}
\usepackage[english]{babel}
\usepackage{amsmath}
\usepackage{amssymb}
\usepackage{graphicx}
\usepackage{float}
\usepackage{titlesec}
\usepackage{soul}
\usepackage{siunitx}
\usepackage{xcolor}
\usepackage{lineno}

\setcitestyle{super,open={},close={}}

\makeatletter
\@ifundefined{textcolor}{}
{%
 \definecolor{BLACK}{gray}{0}
 \definecolor{WHITE}{gray}{1}
 \definecolor{RED}{rgb}{1,0,0}
 \definecolor{GREEN}{rgb}{0,1,0}
 \definecolor{BLUE}{rgb}{0,0,1}
 \definecolor{CYAN}{cmyk}{1,0,0,0}
 \definecolor{MAGENTA}{cmyk}{0,1,0,0}
 \definecolor{YELLOW}{cmyk}{0,0,1,0}
}

\newcommand*{\balancecolsandclearpage}{%
  \close@column@grid
  \clearpage
  \twocolumngrid
}

\@ifundefined{definecolor}
 {\usepackage{color}}{}

\addto\captionsenglish{}

\makeatother

\titlespacing*{\section}
{500pt}{10pt}{0pt}
\titlespacing*{\subsection}
{0pt}{5.5ex plus 1ex minus .2ex}{4.3ex plus .2ex}

\usepackage{dcolumn}
\usepackage{bm}

\usepackage[unicode=true,bookmarks=false,breaklinks=true,pdfborder={0 0 1},backref=false,colorlinks=false]{hyperref} 

\usepackage[
tmargin=1.8cm,
    bmargin=1.8cm,
    lmargin=1.8cm,
    rmargin=1.5cm,
]{geometry}

\makeatletter
\renewcommand\frontmatter@abstractwidth{\dimexpr\textwidth-1in\relax}
\makeatother

\usepackage{lettrine}

\newcommand{\parallelsum}{\mathbin{\!/\mkern-5mu/\!}}

\begin{document}

\title{Superluminal X-waves in a polariton quantum fluid}

\author{Antonio~Gianfrate}
\affiliation{%
CNR NANOTEC, Istituto di Nanotecnologia, Via Monteroni, 73100 Lecce, Italy
}%

\author{Lorenzo~Dominici$^*$}%
\affiliation{%
CNR NANOTEC, Istituto di Nanotecnologia, Via Monteroni, 73100 Lecce, Italy
}%

\author{Oksana~Voronych}
\affiliation{%
Inst.~of Theoretical Physics and Astrophysics, Un.~of Gda\'{n}sk, ul.~Wita Stwosza 57, 80-952 Gda\'{n}sk, Poland
}%

\author{Micha\l{}~Matuszewski}
\affiliation{%
Institute of Physics, Polish Academy of Sciences, Al.~Lotnik\'{o}w 32/46, 02-668 Warsaw, Poland
}%

\author{Magdalena~Stobi\'{n}ska}
\affiliation{%
Faculty of Physics, University of Warsaw, Pasteura 5, 02-093 Warsaw, Poland\\
Correspondence: L Dominici, Email: lorenzo.dominici@gmail.com
}%
\author{Dario~Ballarini}
\author{Milena~De Giorgi}
\author{Giuseppe~Gigli}
\author{Daniele~Sanvitto}
\affiliation{%
CNR NANOTEC, Istituto di Nanotecnologia, Via Monteroni, 73100 Lecce, Italy
}%


\begin{abstract}
\textbf{
In this work we experimentally demonstrate for the first time spontaneous generation of two-dimensional exciton-polariton X-waves. X-waves belong to the family of localized packets, which are capable of sustaining  their shape with no spreading even in the linear regime. This allows to keep the packet shape and size for very low densities and very long times compared, for instance, to soliton waves, which always necessitate a nonlinearity to compensate the diffusion.
Here we exploit the polariton nonlinearity and unique structured dispersion, comprising both positive- and negative-mass curvatures,
to trigger an asymmetric four wave mixing in the momentum space. 
This ultimately enables self-formation of a spatial X-wave front.
By means of ultrafast imaging experiments we observe the early reshaping of the initial Gaussian packet into the X-pulse and its propagation even for vanishing small densities.
This allows us to outline the crucial effects and parameters driving the phenomena and to tune the degree
of peak superluminal propagation, which we found to be in a good agreement with numerical simulations.\\
}
\end{abstract}

\maketitle



\noindent \textbf{INTRODUCTION}\\
\lettrine{X}{-waves} (XWs)\cite{hernandez-figueroa_localized_2008,recami_chapter_2009} are a specific type of nonspreading wavepackets, which maintain their transverse shape along a large field-depth with respect, e.g., to Gaussian beams or packets.
Another well known class of nonspreading waves, are solitons. However, in case of solitons, dispersion is constantly compensated by nonlinearity in the medium. XWs, instead, as other types of nonspreading waves, generally formed by Bessel beams, can maintain their shape also in absence of nonlinearity \cite{sonajalg_demonstration_1997}.
They represent a high-interdisciplinary topic, spanning from photonics to acoustics and are relevant in any system which is governed by the wave equation.
The first experimental demonstration of such optical waves employed a cw laser light \cite{durnin_diffraction-free_1987}.
These beams are not free from diffraction, but their transverse profile is such that keeps its main peak well confined, while the weaker lateral ones expand upon propagation. The same \textit{localization} principle holds for pulsed XW packets
that, in essence, are a polychromatic superposition
of Bessel beams \cite{salo_unified_2000}.

Since the early 90s XWs have been extensively studied both theoretically and experimentally using nonspreading acoustic pulses by Lu and Greenleaf \cite{lu_ultrasonic_1990,lu_nondiffracting_1992}. Later XWs were obtained with light by injecting sub-ps laser pulses across a dispersive material \cite{sonajalg_demonstration_1997}, demonstrating the potential for signal transmission and imaging. Indeed their applications span 
different systems\cite{hernandez-figueroa_localized_2008}, from medical ultrasound scanning to optical coherence tomography,
nondestructive evaluation of materials and defects identification; 
from free space optical and radio-based telecommunication systems
to optical tweezers, such as accelerating or guiding beams, 
also within plasmonics near-field manipulation, microscopy and signal transmission.
In nanotechnology, the localized waves allow to reliably produce high quality beams, 
required for optical and electron-beam lithography with sub-diffraction resolution \cite{yalizay_fabrication_2012}.
Any of these experimental cases deals with, e.g., realistic antennas truncated in time and space,
and analytical solutions have been found too for the finite energy content cases \cite{zamboni-rached_new_2002,zamboni-rached_superluminal_2003}.
Among their fascinating properties, it's worth mentioning that X-waves in vacuum also correspond to the simplest superluminal waveforms \cite{zamboni-rached_new_2002,zamboni-rached_superluminal_2003}, i.e., solutions with effective velocity higher than $c$, which emerges from the superposition of ordinary Bessel beams \cite{durnin_diffraction-free_1987,salo_unified_2000} and at the same time conform with the constraints of special relativity and the causality principle \cite{hernandez-figueroa_localized_2008}.

\begin{figure*}[htbp]
\includegraphics[width=0.9\textwidth]{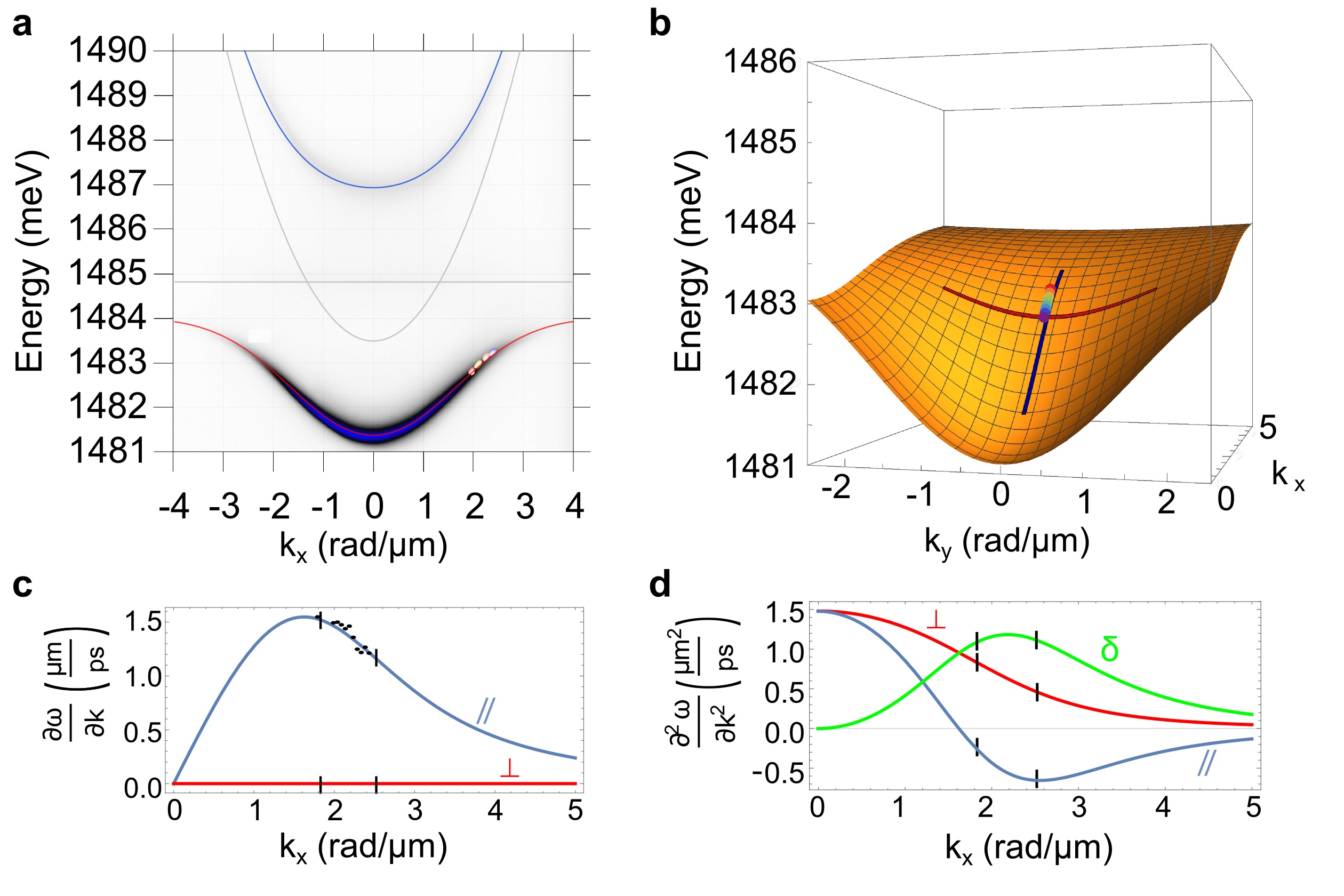}
\caption{\textbf{Polariton dispersion.} \textbf{a}, Experimental dispersion under non-resonant pumping. The superimposed points (different colors) on the right side of the lower branch represent polariton emission under resonant excitation at different momenta $k$. The solid lines are theoretical fits.
\textbf{b}, 3D representation of the $E(k_x, k_y)$ dispersion surface around the inflection point, highlighting the central longitudinal crosscut (blue line) and an exemplificative transverse one (red line).
\textbf{c}, First derivative (group velocity) of the dispersion along the propagation direction ($\parallelsum$, blue) and the transverse one ($\perp$, red).
The black dots represent the experimental group velocities (retrieved by independent experiments on the fluid dynamics). \textbf{d}, Second derivative (inverse effective mass) of the dispersion along the propagation direction ($\parallelsum$, blue) and the transverse one ($\perp$, red), and their difference ($\delta$, green). The vertical black lines define the explored $k$ region which is maximizing the effective mass anisotropy.}
\label{disp}
\end{figure*}

The renewed interest in the XWs is also driven by their potential applicability in the field of atomic Bose-Einstein condensates (BEC) \cite{conti_nonspreading_2004} and dissipative polariton condensates \cite{voronych_exciton-polariton_2016}. Both these systems bear deep similarity to the electromagnetic case because all these systems can be described by nonlinear Schr\"{o}dinger equations \cite{efremidis_exact_2009,conti_generation_2004}. 
Polariton XW solutions were predicted not only for microcavity polaritons \cite{voronych_exciton-polariton_2016}, 
but also in the case of Bragg polaritons, periodically embedding quantum wells directly into the multilayer stacks. \cite{sedov_hyperbolic_2015}
In both cases
the XWs solutions rely on the locally hyperbolic dispersion (i.e., including both negative and positive curvatures).
Several theoretical proposals have been developed on the topic as well as on the possibility to obtain spontaneous X-wave upon exploitation of the nonlinearities \cite{couairon_nonlinear_2006,kolesik_dynamic_2004,conti_nonlinear_2003,di_trapani_spontaneously_2003}.
Recently, a quantum description of XWs has been also developed, highlighting the difference in the entanglement properties between externally imprinted and spontaneously generated states \cite{ciattoni_quantum_2007}. However X-waves have not been imprinted or generated in a BEC yet.

Our work bridges the X-wave idea to hybrid fluids of light and matter.
We report for the first time on the experimental self-generation of an XW packet in a two-dimensional exciton-polariton superfluid, starting from an initial Gaussian photonic pulse. This effect has been achieved upon fine tuning of the polariton nonlinearity and proper balance of the positive/negative effective masses ratio along the transverse/longitudinal directions, respectively. 
Using ultrafast digital holography, the experiments show the initial pulse reshaping and propagation, demonstrating its longitudinal localization down to vanishing densities of the packet. 
Noteworthy, the 2D polariton geometry allows to obtain the axial XW density and phase profiles along the propagation direction.
The optical access to the wavefunction phase allows to highlight some peculiar topological defects associated to the specific way we obtain the X-wave.  Moreover, upon changing uniquely the initial nonlinearity amount, we show a tunable degree of superluminal peak speed, with respect to the group velocity of the polariton system. 

Microcavity exciton polaritons~\cite{sanvitto_road_2016,byrnes_excitonpolariton_2014,dagvadorj_nonequilibrium_2015,deng_exciton-polariton_2010,amo_collective_2009,kasprzak_boseeinstein_2006,balili_bose-einstein_2007} are bosonic particles which result from mixing of two quasi-parabolic modes, the QW (quantum well) excitons and the MC (microcavity) photons, with dispersions of highly unbalanced curvatures. The anticrossing feature of the bare modes, associated to the strong coupling regime, finally produces a highly non-parabolic shape of one of the new normal modes, namely the lower polariton branch (LPB) \cite{kavokin_microcavities_2017}. In particular the presence of an inflection point, representing both a maximum of the group velocity ($v_g= \partial \omega /\partial k$) and an inversion of the so called \textit{diffusive} effective mass [$m_{diff}= (\partial^2 \omega /\partial k^2 )^{-1}$] \cite{colas_self-interfering_2016}, is the fundamental reason for the XW spontaneous formation.
Polaritons also exhibit very strong nonlinearities \cite{walker_ultra-low-power_2015,vladimirova_polariton-polariton_2010} able to achieve superfluid regimes \cite{amo_superfluidity_2009,berceanu_multicomponent_2015} supporting quantized vortex \cite{amo_polariton_2011,sanvitto_persistent_2010}, or leading to several pattern  \cite{whittaker_polariton_2017,dominici_real-space_2015,manni_spontaneous_2011,wertz_spontaneous_2010} and soliton states formation \cite{ostrovskaya_dissipative_2012,sich_observation_2012}.
However, we note that the XW, a solution which exists in the linear limit, is fundamentally different from the two-dimensional bright solitons discussed in \cite{sich_observation_2012,egorov_two-dimensional_2010}, where localization was achieved in the so called bistability regime, in which the soliton wavepacket was supported by an additional background pump. 
While solitonic wavepackets are well suited for polaritonic devices
that utilize nonlinearity, such as logic gates or transistors \cite{ballarini_all-optical_2013},
they are inherently fragile against particle loss that is unavoidable
in any photonic system. On the other hand, linear localized solutions
are fundamentally robust against losses, and have a potential for
applications in data transport between distant system components. This
approach has been demonstrated to be efficient in overcoming
performance bottleneck in electronic signal processing \cite{sun_single-chip_2015}.\\


\noindent \textbf{MATERIALS AND METHODS}\\
\noindent \textbf{Experimental methods}\\
\noindent 
The experiments described here are performed on a GaAs/Al$_x$GaAs microcavity (MC) composed by three quantum wells (QW) enclosed by two distributed Bragg reflectors,
the details of which can be found in Ref.~\cite{colas_polarization_2015,dominici_ultrafast_2014}. 
The positions inside the MC are set in order to have the QWs in the antinodes of the confined photonic field.
The strong coupling of the two bare modes, the photonic ($\psi_C$) and excitonic ($\psi_X$) fields, lead to two new hybrid modes, the LPB and the upper polariton branch (UPB).
This sample is also grown on a specific doped GaAs substrate with a transparency window centered at $830~\text{nm}$, that consequently allows to work in a transmission configuration. The sample is kept at a constant temperature of $10~\text{K}$ by means of a cryostat avoiding thermal ionization of excitons.

The setup in use during this experiment is an ultrafast digital holography setup described in detail in Ref.~\cite{colas_polarization_2015,dominici_ultrafast_2014}, where the emission signal is let interfere with a homodyne uniform plane wave reference. The two beams are sent with a slightly different incidence angle on a charge coupled device camera to collect the associated interference pattern.
The resulting interferograms are analyzed by a digital Fast Fourier Transformation to obtain the amplitude ($\psi$) and the phase ($\varphi$) of the complex wavefunction in real space. A delay line on the reference optical path allows us to scan the signal in time upon changing the time delay of the reference.
The temporal resolution of this technique is mainly limited by the duration of the picosecond laser pulse in use, $2.5~\text{ps}$, which in the current experiments allows to selectively excite the lower polariton mode upon proper tuning at $\sim 836~\text{nm}$. The time step was set to $0.5~\text{ps}$. 
 
Circular polarization is set in the excitation beam in order to generate only one spin population and consequently maximize the interactions. The same reshaping effects are obtained upon a double total population density when using a linearly polarized excitation beam. The pump spot is set to 
a $\text{FWHM}_{x,y} = 10~\mu\text{m}$ ($\text{FWHM}_{kx,ky} = 0.6~\mu \text{m}^{-1}$ in the reciprocal space), in order to facilitate the nonlinear scattering process in real space and have a wide enough spot to cover the interested dispersion range in $k$ space.\\

\noindent \textbf{Numerical methods}\\
\noindent
To illustrate the dynamical X-wave formation and localization induced by the nonlinearity, we performed simulations starting from a Gaussian initial state $\Psi(x,y) = \frac{1}{2\pi\sigma^2}\exp\{-(x^2+y^2)/(2\sigma^2)\}$ where $\sigma=\text{FWHM}/(2\sqrt{2\ln 2})$ and FWHM is the full width at half-maximum of the Gaussian spot. The GPE described in the text was solved numerically using the Runge--Kutta method of 4th order. 
The device parameters were: $m_C = 4.27 \times 10^{-5} ~m_e$,
$\gamma_C = 0.2~\text{ps}^{-1}$, $\gamma_X = 0.2~\text{ps}^{-1}$, $g = 2 \times
10^{-3}~\text{meV} \cdot \mu\text{m}^2$, photon-exciton detuning $\Delta = -0.55~\text{meV}$, $\Omega_R = 5.4~\text{meV}$. The details of the numerical method are described in~\cite{voronych_numerical_2017}. Numerical computations were performed with Zeus cluster of the ACK ``Cyfronet'' AGH computer center.\\

\noindent \textbf{RESULTS AND DISCUSSION}\\
\noindent \textbf{Dispersion and effective masses}\\
\noindent
The high quality factor Q of our microcavity converts in that the LPB and UPB modes are  well separated with respect to their linewidths.
The two modes can hence manifest their dispersions, observed upon collecting the non-resonantly excited fluorescence as shown in Fig.~\ref{disp}(a). Here, we will focus our attention only on the LPB which shows a strong non-parabolic behavior at higher $k$ vectors. Further experimental details can be found in the Methods.
The 3D representation of the LPB dispersion surface $E(k_x, k_y)$ is shown in Fig.~\ref{disp}(b) in a region around the inflection point ($k_x \sim 1.62~\mu \text{m}^{-1}$). In the figure we highlight the non-parabolic character by reporting two orthogonal crosscuts along the longitudinal direction ( $\parallelsum$ blue curve, centered at $k_y = 0$) and along the transverse direction ($\perp$ red curve, at $k_x = 2.15~\mu \text{m}^{-1}$). The noteworthy feature that can be appreciated from the 3D representation is the opposite curvature of the two slices around the inflection point.

\begin{figure}
\includegraphics[width=0.98\columnwidth]{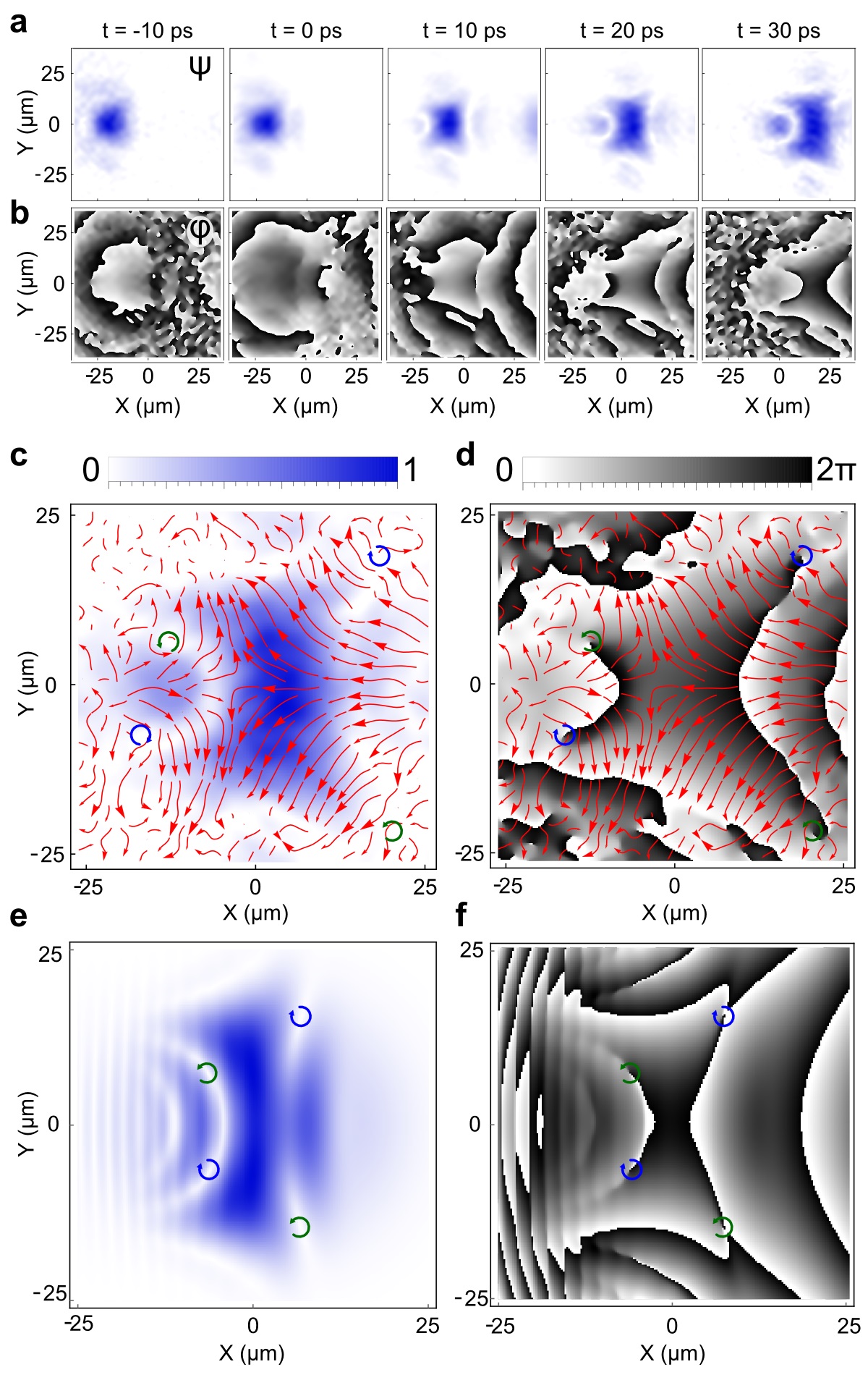}
\caption{\textbf{Dynamical XW sequence.}
\textbf{a},\textbf{b}, Spatial distribution of the polariton wavefunction amplitude (\textbf{a}) and phase (\textbf{b}) at different time frames ($t = -10$, $ 0$, $ 10$, $ 20$ and $30~\text{ps}$). $\text{t}=0$ corresponds to the end of pulse injection and consequently to the start of the free evolution.
The phase map is represented with respect to the moving-packet's frame of reference (see also Movie 1 and 2).
\textbf{c}, \textbf{d} Detailed maps of the amplitude (\textbf{c}) and phase (\textbf{d}) ($t = 27.5~\text{ps}$), with superposition of the relative in-plane momenta distribution (red arrows) and of the quantized vortices (blue and green circles).
\textbf{e}, \textbf{f}, GPE simulated  amplitude (\textbf{e}) and phase (\textbf{f}) maps 
at $27.5~\text{ps}$ for a $10~\mu\text{m}$ wide pump spot and a wavevector of $2.55~\text{$\mu$m$^{-1}$}$.
All the amplitude maps are normalized to their maximum at the specific time frame.
}
\label{opt.eff.}
\end{figure}

Moving along the central longitudinal line, the dispersion geometry always corresponds to a null transverse velocity [$\partial \omega/\partial k_y(k_x,k_y=0$)]. The longitudinal group velocity instead grows till a maximum ($1.5~\mu\text{m/ps}$) at the inflection point and decreases  for larger in-plane longitudinal momenta ($k_x$), Fig.~\ref{disp}(c). 
At the same time, both the longitudinal and transverse curvatures of the dispersion surface change as a function of $k_x$, as clearly illustrated in Fig.~\ref{disp}(d).
In particular, the curvatures have opposite sign inside the investigated region (the explored range is denoted by  dots or by vertical ticks in any of the four panels), and this corresponds to opposite effective masses.\\

\noindent \textbf{Polariton X-wave}\\
\noindent
We resonantly excite the polariton superfluid by means of $2.5~\text{ps}$ laser pulses tuned at $\sim 836~\text{nm}$ and focused to a $\sim 10~\mu \text{m}$ diameter spot.
In Fig.~\ref{opt.eff.} we show the dynamics of the effect experimentally optimized using $k = 2.35~\mu \text{m}^{-1}$ and $75~\mu \text{W}$ pumping power. Figure \ref{opt.eff.}(a) displays the modulus and Fig.~\ref{opt.eff.}(b) the phase of the polariton X-wavepacket. The time zero in the temporal evolution is set when the pump stops injecting polaritons and these are left free to evolve within their lifetime.
Initially, the density distribution reveals a Gaussian shape with a rather homogeneous phase (with just a weak radial gradient associated to the beam curvature). However, after $10~\text{ps}$ the X-shape can be already neatly distinguished. At the successive $t = 20~\text{ps}$ and $t = 30~\text{ps}$ snapshots, we can appreciate just a small vertical spread of the packet however, without a significant distortion in the shape. Noticeably, the longitudinal waist size remains essentially constant, despite the polariton lifetime being as short as $\sim 10~\text{ps}$~\cite{colas_polarization_2015,dominici_ultrafast_2014}.

A very interesting feature can be spotted in the phase map, as shown in Fig.~\ref{opt.eff.}(b): the appearance of four quantized vortices, at the edges of the packet. They are shown in detail in the maps of Fig.~\ref{opt.eff.}(c,d), overlapping the streamlines of the phase gradient (red arrows) and the dots of the phase singularities (blue and green arrow circles). 
The diagonally-displaced vortex-antivortex pairs are an expression of the hyperbolic topology of the driving in-plane momenta. 
Indeed, as evident in the centre of the packet, the flows are pushing the polaritons inwards along the propagation direction, consequently keeping the signal compact, and outwards in the transverse direction.

\begin{figure*}[t]
\includegraphics[width=0.9\textwidth]{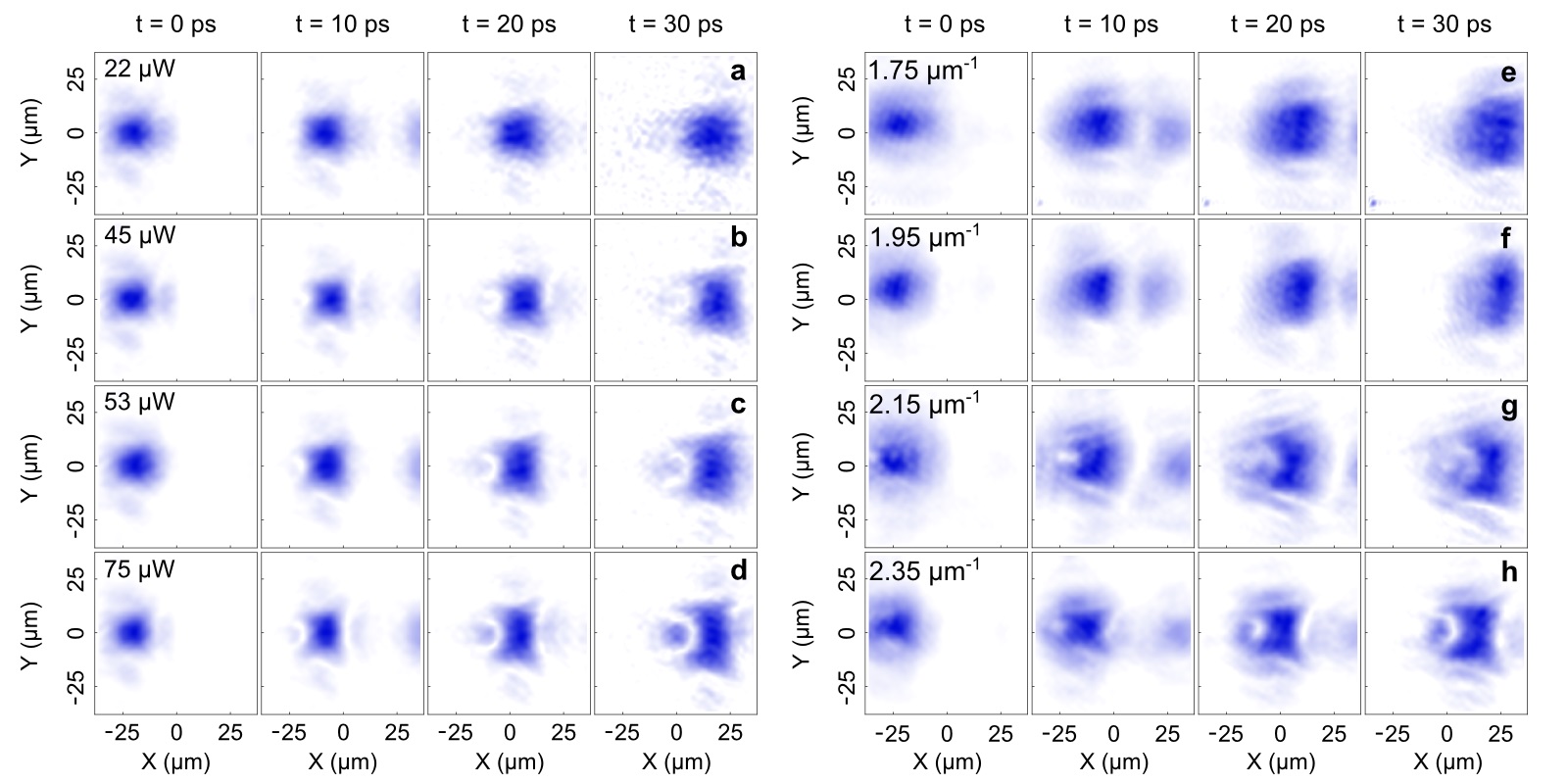}
\caption{\textbf{Tuning of the packet reshaping.} \textbf{a},\textbf{b},\textbf{c},\textbf{d}, Each row corresponds to a different power and the optimum injection wavevector ($k_x= 2.35~\mu \text{m}$). \textbf{e},\textbf{f},\textbf{g},\textbf{h}, The pumping power is constant ($80~\mu \text{W}$), while the in plane momentum varies. The secondary spot appearing in front of the main signal at $\sim 10~\text{ps}$ is a reflection from the substrate edge. Their separation is increasing with the angle of injection (i.e., with $k$). Such reflection is not affecting the main packet especially in the case of the $k$ used for the optimized X-wave (separation $\sim$ two times the spot width). }
\label{comp.pann.pow}
\end{figure*}

The dynamics of the polariton superfluid was succesfully modeled within the mean field approximation, by a set of coupled equations equivalent to the Gross-Pitaevskii equation (GPE):
\begin{equation}
  \label{eq:full_X}
  \kern-0.5em\left\{\kern-0.25em\begin{aligned}[c]
  i\hbar\dfrac{d}{dt}\psi_C = \frac{\Omega_R}{2}\psi_X
  + \Big(\dfrac{\hbar\gamma_C}{2i} - \dfrac{\hbar^2}{2m_{C}}\partial^2_x
  - \dfrac{\hbar^2}{2m_{C}}\partial^2_y\Big) \psi_C ,\\
  i\hbar\dfrac{d}{dt}\psi_X = \frac{\Omega_R}{2}\psi_C + \big(\dfrac{\hbar\gamma_X}{2i}+g\lvert\psi_X\rvert^2\big)\psi_X,
  \end{aligned}\kern-1em\right.
\end{equation}
where $m_{C}$ is the effective mass of microcavity photons, $\Omega_R$ is the Rabi frequency coupling the photonic $\psi_C$ and excitonic $\psi_X$ fields, $\gamma_C$ and $\gamma_X$ are the associated decay rates and $g$ is the nonlinear interaction term in the exciton component. 
Further details are given in Methods.
The results shown in Fig.~\ref{opt.eff.}(e,f) represent the amplitude and phase maps, respectively, at a time of $27.5~\text{ps}$, demonstrating a very good agreement with the main experimental features. The modulation in the tail of the signal that can be spotted in the theoretical Fig.~\ref{opt.eff.}(e,f) could be due to the interference with a weak nonlinear scattering to opposite $k_x$ states. This may be not visible in the experimental data due to the temporal resolution ($2.5~\text{ps}$ limited by the reference pulse).

The opposite transverse and longitudinal effective masses confer to the GPE---that describes the polariton dynamics---a highly hyperbolic character. This behavior is crucial to sustain the X-wave phenomena, and demonstrates that the shape conservation does not rely on the nonlinearity as in the case of solitons \cite{dominici_real-space_2015,sich_observation_2012}, but rather on the dispersion morphology.
It was previously shown that an X-shaped initial profile can be a stationary solution of the linear GPE model \cite{voronych_exciton-polariton_2016}. We experimentally demonstrate that in a weakly nonlinear regime, an initial Gaussian state can be triggered to spontaneously evolve into a steady X-wave via an early FWM (Four Wave Mixing) process.\\

\noindent \textbf{Nonlinear triggering}\\
\noindent
Although the nonlinearities play no role in the propagation and sustain of the signal, they are crucial for the initial reshaping of the Gaussian pulse into the X-packet. 
Indeed, the choice of the initial spot size in the real space (density FWHM $\sim 10~\mu \text{m}$) enables to achieve a proper extension in the reciprocal (momentum) space (FWHM $\sim 0.6~\mu \text{m}^{-1}$), and thus, to exploit the negative curvature.
The nonlinearity allows an asymmetric reshaping in the momentum space, according to the dispersion shape.
As a result of that, an elongated spot in the reciprocal $k$ space is created along the direction of propagation, which is a signature of a stronger confinement in the real space.

To highlight the impact of nonlinearities the temporal dynamics at four different pumping powers are shown in Fig.~\ref{comp.pann.pow}. At low density, Fig.~\ref{comp.pann.pow}(a), the reshaping is absent and the signal spreads uniformly in both directions, longitudinal and transverse to the propagation.
However, an anisotropy in intensity distribution between the longitudinal and transverse diffusion starts to appear for increasing pump power, as is shown in Fig.~\ref{comp.pann.pow}(b,c). 
At $75~\mu \text{W}$  the reshaping reaches its optimum, Fig.~\ref{comp.pann.pow}(d), and the packet shows a very well defined X-shape, together with a small circular tail. Above this power the dynamics enter a strongly nonlinear regime (between $100~\mu \text{W}$ and $500~\mu \text{W}$), where the redistribution due to high densities involves radial counterflows which reshape the signal beyond a recognizable X-packet. Such regime is just before the onset of the dynamical nonlinearity inversion leading to the real space collapse described in \cite{dominici_real-space_2015}.

The role played by the in-plane momentum $k_{x}$ is shown in Fig.~\ref{comp.pann.pow}(e-h) where only the injection angle is changed, while keeping constant the initial density.
In Fig.~\ref{comp.pann.pow}(e), despite that the nonlinearities are as high as in Fig.~\ref{comp.pann.pow}(d), no redistribution is observed. Upon gradually increasing the injection angle, Fig.~\ref{comp.pann.pow}(f) and (g), the packet shows again a marked anisotropy in the diffusion along the longitudinal and transverse direction. 
This is due to the larger difference between the longitudinal and transverse effective masses in the excited region of the dispersion. This difference reaches its maximum at $2.35~\mu \text{m}^{-1}$ where the reshaping is optimized, Fig.~\ref{comp.pann.pow}(h).\\

\begin{figure}[h]
\includegraphics[width=1.0\columnwidth]{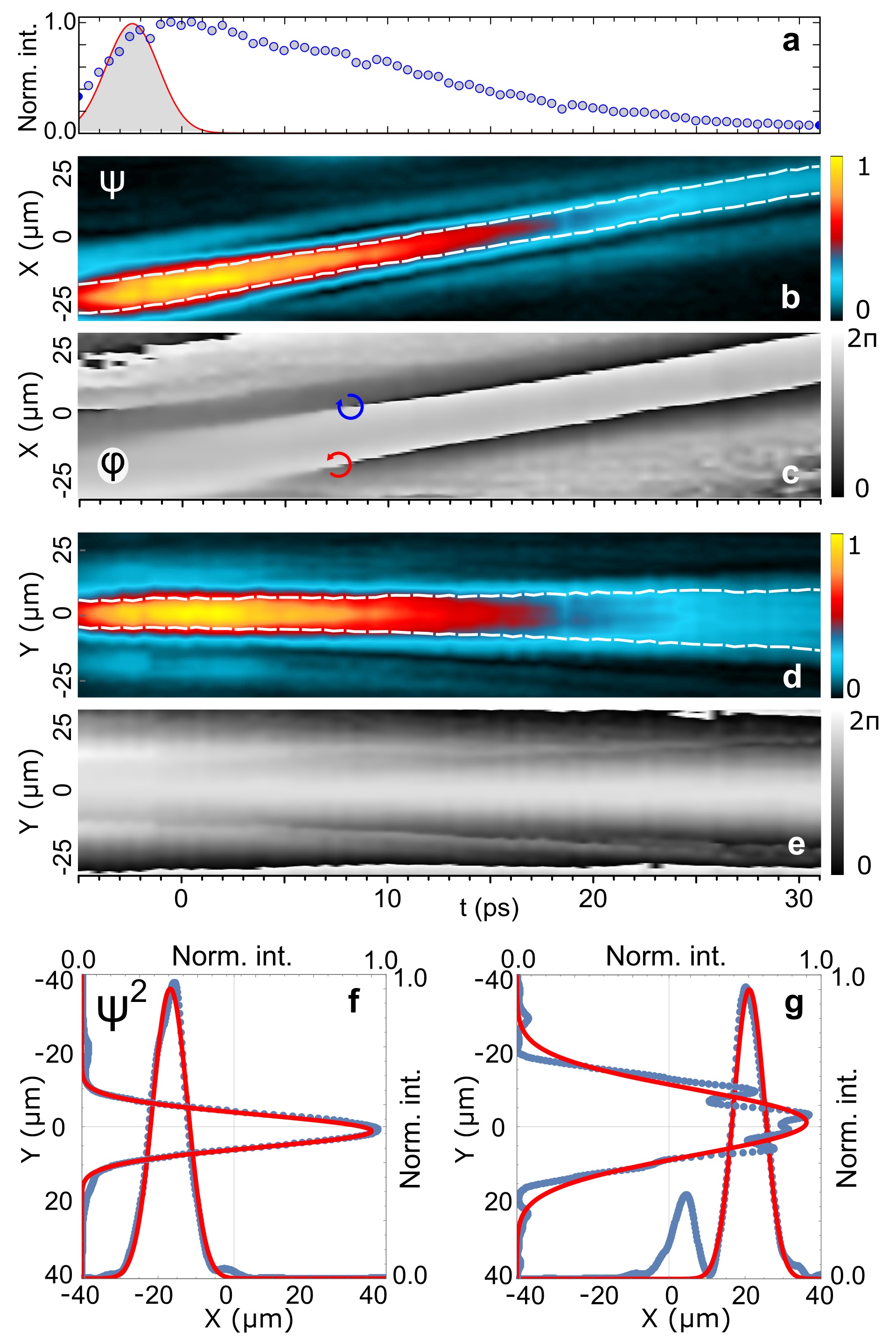}
\caption{
\textbf{Timespace profiles of the X-wave.}
\textbf{a}, In blue, time evolution of the normalized total polariton population, in red the temporal envelope of the pump pulse.
\textbf{b},\textbf{c},\textbf{d},\textbf{e}, Vertical (\textbf{b},\textbf{c}) and horizontal (\textbf{d,\textbf{e}}) timespace charts of the amplitude and phase. 
The dashed white lines in \textbf{b} and \textbf{d} represent the temporal evolution of the FWHM of the packet density. 
In \textbf{c} the longitudinal phase is shown with respect to the moving frame of reference.
Moreover, 
in both \textbf{c},\textbf{e} the phase is pinned to be constant in the center of the main packet.
The circled arrows in \textbf{c} represent two phase singularities in the time-space domain.
In \textbf{d},\textbf{e} the transverse crosscuts are taken at a longitudinal position which follows the packet maximum.
\textbf{f},\textbf{g}, Longitudinal and transverse normalized density profiles at $t = 0~\text{ps}$ (\textbf{f}) and $t = 30~\text{ps}$ (\textbf{g}). Solid lines are Gaussian fits. 
The dynamics are relative to $P = 75~\mu\text{W}$ and $k_{x}=2.35~\mu \text{m}^{-1}$ [as in Fig.~\ref{comp.pann.pow}(d)].
}
\label{chart}
\end{figure}

\noindent \textbf{Localization and superluminality}\\
\noindent
We now focus on the propagation of the polariton XWs. Figure~\ref{chart}(a) shows the time evolution of the normalized polariton population (blue points) together with the pump pulse temporal envelope (solid red curve). The $t = 0~\text{ps}$ has been chosen at the maximum of the polariton population, when the pulse has essentially finished its pumping action and the polaritons free evolution starts.
The growing longitudinal/transverse anisotropy can be appreciated upon a visual comparison between the associated amplitude timespace charts in Fig.~\ref{chart}(b) (longitudinal) and Fig.~\ref{chart}(d) (transverse), as well as from the associated phase charts in Fig.~\ref{chart}(c) and Fig.~\ref{chart}(e). 
In the longitudinal charts the signal propagates for $40~\mu\text{m}$ with a constant speed of $\sim 1.20~\mu\text{m}/\text{ps}$, with a final width very much the same as the original shape, while in the transverse maps, the width reveals the standard wave-packet diffusion.
This is clearly confirmed in Fig.~\ref{chart}(g,h), where both the longitudinal and transverse profiles are reported for $t= 0$ and $t= 30~\text{ps}$, respectively, together with the associated Gaussian fits.

\begin{figure*}[t]
\includegraphics[width=0.85\textwidth]{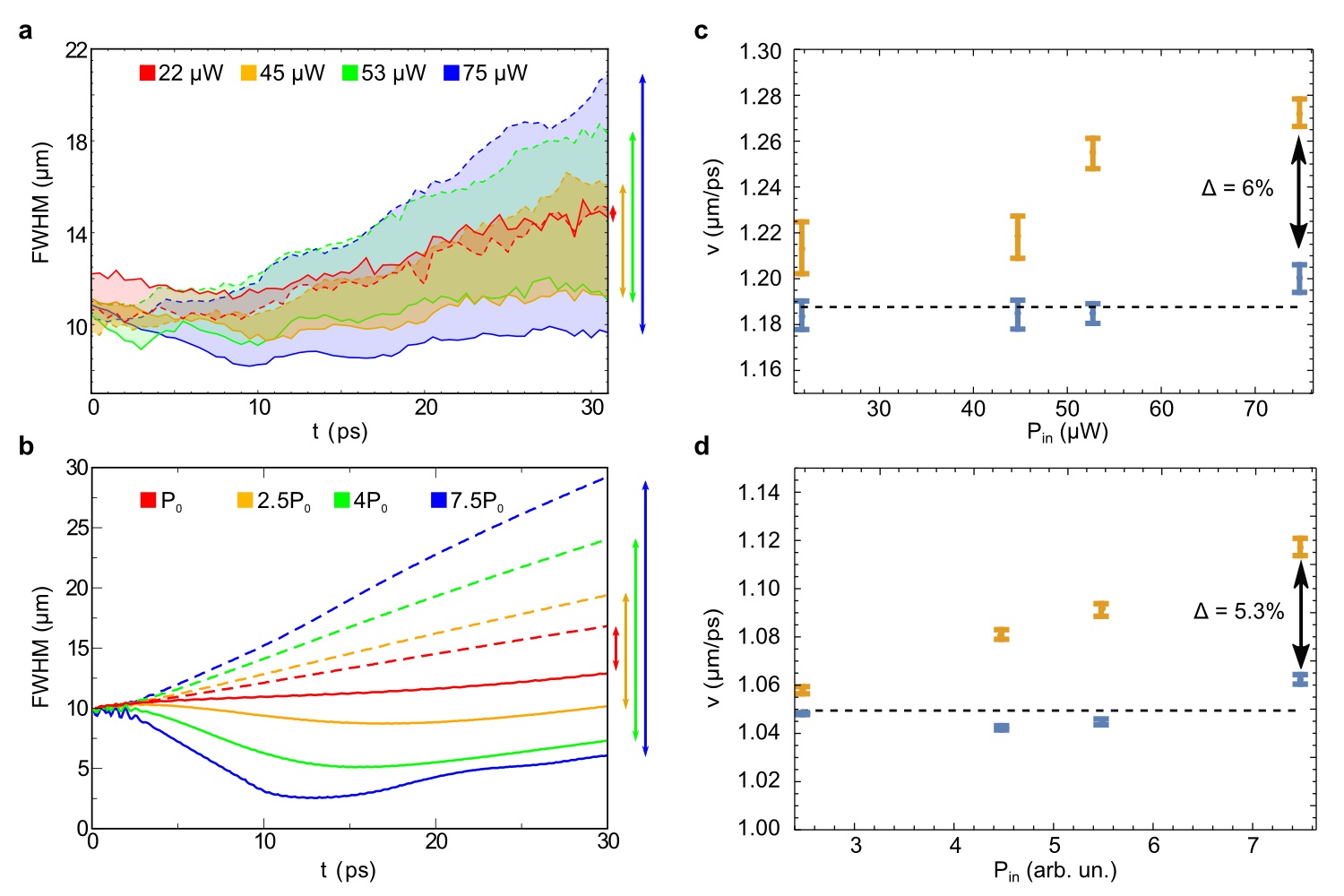}
\caption{
\textbf{Wavepacket spreadings and speeds.}
\textbf{a}, Experimental longitudinal (solid line) and transverse (dashed line) density FWHMs during time at four different excitation powers for $k_{x} = 2.35~\mu\text{m}^{-1}$.
\textbf{b}, Simulation of the time evolution of the longitudinal (solid line) and transverse (dashed line) density FWHMs, for different initial total populations and $k_{x}= 2.55~\mu\text{m}^{-1}$ in-plane momentum. The arrows on the right side in both panels indicate the difference between the longitudinal and transverse widths.
\textbf{c}, Experimental velocities of the center-of-mass (blue dots) and peak (orange dots) positions at four different excitation powers (at $k_{x} = 2.35~\mu\text{m}^{-1}$).
\textbf{d}, Simulated velocities of the center-of-mass (blue dots) and peak (orange dots) positions at four excitation powers (at $k_{x} = 2.55~\mu\text{m}^{-1}$).
The arrows on the right side in both panels indicate the percentage difference between the peak and center-of-mass speeds at the largest power.
The horizontal dashed lines represent the average in the center-of-mass velocities. The different values between experimental and computational values are due to the different used $k_x$ and  to a not perfect calibration of the simulation parameters.
}
\label{width}
\end{figure*}

The power dependence of the differential spreading along the two directions is analyzed in detail in Fig.~\ref{width}(a). 
Here the temporal evolution of the longitudinal and transverse density FWHMs is shown for different excitation powers, corresponding to the previous Fig.~\ref{comp.pann.pow}(a-d).
For the lowest power ($P=22~\mu\text{W}$, red line), the reshaping in completely absent and the wavepacket expands continuously in both directions, longitudinal (filled dots) and transverse (open dots) with the same spreading rate.
At larger injected power, and consequently stronger nonlinearity, the degree of anisotropy between the longitudinal and transverse size gradually increases ($P=45~\mu\text{W}$, $P=53~\mu\text{W}$, orange and green dots, respectively), leading to the suppression of longitudinal spread.
Strikingly, for $P=75~\mu\text{W}$, 
the packet even undergoes a longitudinal squeezing during the first $10~\text{ps}$, 
associated to the nonlinear redistribution into the X-wave packet, 
 whose shape can be neatly distinguished in the previous maps of Fig.~\ref{comp.pann.pow}(d).
Based on these features we may state that it is possible to qualitatively distinguish three main different phases in the dynamics:~pulse injection ($-5 \div 0~\text{ps}$), initial redistribution ($0 \div 10~\text{ps}$), propagation ($10 \div 30~\text{ps}$).
The numerical simulations of Fig.~\ref{width}(b) reproduce the experiments in a perfect agreement with our trends.
We remark again that this phenomenology is different from the bright solitons which are sustained under a cw background pump beam as in \cite{sich_observation_2012,egorov_two-dimensional_2010}.
In that case the pump keeps the background just below the bistability threshold all over its width, and feeds the nonlinear sustain of the moving soliton,
which can propagate only within the pump spot.
Instead here, while the transverse width is not conserved (in agreement with the lateral positive mass), the longitudinal width is preserved along a propagation length of more than $40~\mu\text{m}$, despite the packet arrives to such position with only a very small fraction of the initial population and density (less than one order of magnitude lower).
Furthermore, 
the brigth solitons in \cite{sich_observation_2012,egorov_two-dimensional_2010}, 
exhibit a propagation speed which is set uniquely by the injection $k$ of the cw background pump and---being dissipative solitons---it is not affected by the seed pulse.
On the contrary here,
the X-wave offers the possibility 
to tune the group velocity of the packet by the incident in-plane momentum,
and can furthermore achieve 
a fine degree of tunability of the peak speed, 
upon the power control of the exciting pulse.

Indeed, our wavepacket exhibits one of the most interesting signatures of Bessel X-pulses, \emph{superluminality}.
This effect is driven by the Bessel cone angle $\theta$ associated to X-pulses~\cite{bonaretti_spatiotemporal_2009,mugnai_observation_2000,bowlan_measuring_2009,valtna-lukner_direct_2009}, 
whose peak moves (in vacuum) at $v = c/\cos(\theta)$.
In any system, the role of $c$ is played by the group velocity $v_g$ as obtained from the specific dispersion slope
(discussed for polariton waves in Fig.~\ref{disp}). 
Here we experimentally observe an increase in the speed of the density peak with respect to the center-of-mass speed, up to a value of 6\% in the case of the largest power,
as shown in Fig.~\ref{width}(c). 
We can evaluate that a maximum angle $\theta \sim 18^{\circ}$ is reached for $P=75~\mu\text{W}$.
In terms of transverse in-plane momentum such angle corresponds to a $\delta k_y \sim \pm 0.6~\mu\text{m}^{-1}$.
These lateral $k$-states are induced in the initial FWM along the (almost flat) transverse direction of the dispersion.
Numerical simulations performed at different initial population confirm very well the trend in the increase of the peak velocity
with respect to the center-of-mass one, as shown in Fig.~\ref{width}(d).
We would like to stress that different degrees of superluminal speed could here be achieved  
without changing any other parameter
(e.g., spot width in real/$k$-space, central momentum, central energy, pulse width)
but only the pulse power, consequently tuning the strength of nonlinearity.

\begin{figure}[t]
\includegraphics[width=1\columnwidth]{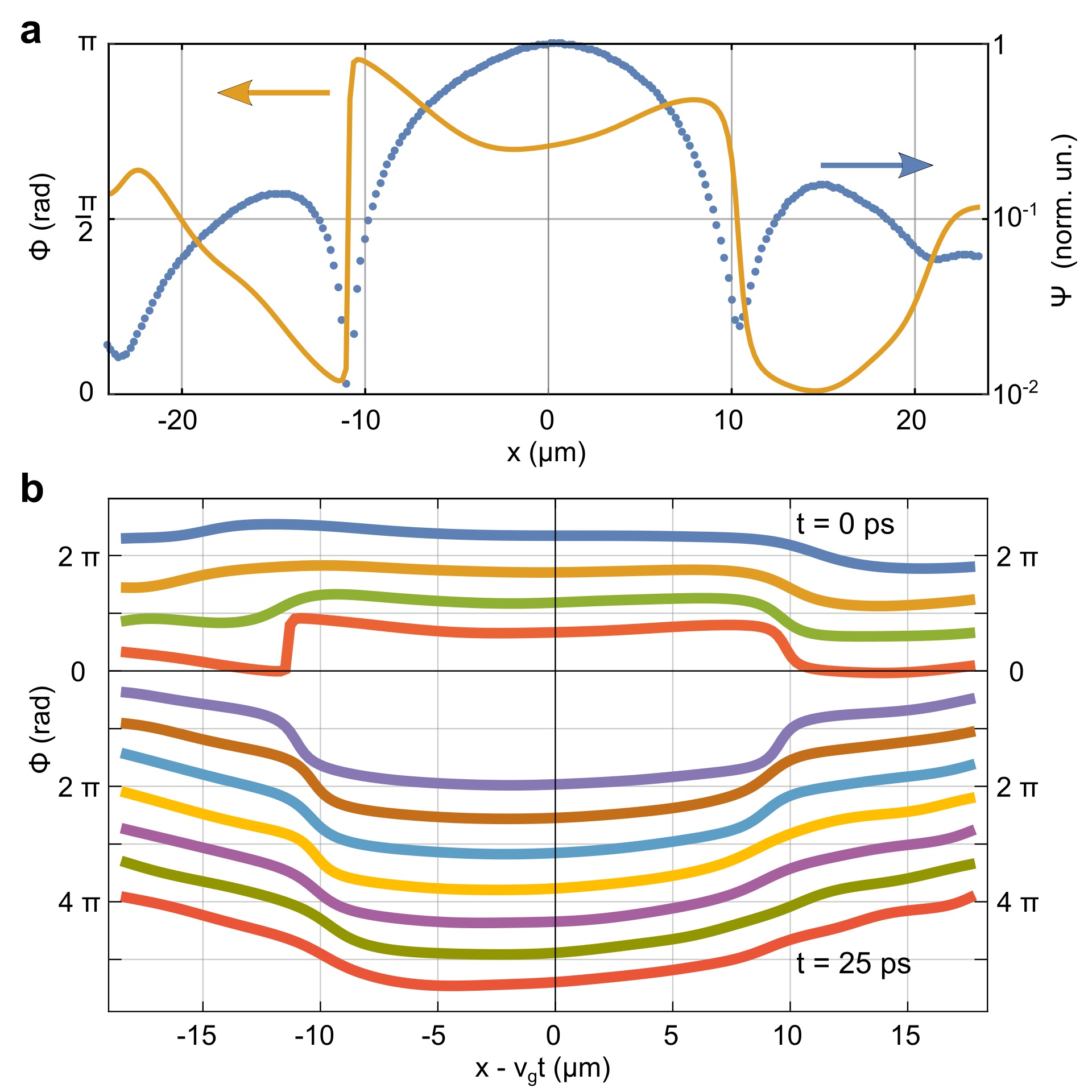}
\caption{
\textbf{Transient dark soliton.}
\textbf{a}, Phase (solid orange line) and amplitude (blue dots) central crosscut profiles along the propagation direction, at $t = 7.5~\text{ps}$ and for $P=75~\mu\text{W}$. 
\textbf{b}, Phase profiles along the longitudinal axis at fixed time intervals of $2.5~\text{ps}$. The top line represents $t = 0~\text{ps}$ while the subsequent profiles are mutually shifted downwards of the same amount for the sake of clarity. The phase profiles are shown 
with respect to the moving frame of reference, and the phase is pinned to be constant at the edge of the packet. 
}
\label{dark.sol.}
\end{figure}

A complementary nonlinear effect is obtained along the longitudinal direction.
As introduced in Fig.~\ref{opt.eff.}, a specific feature of our structured polariton XW is represented by the leading and trailing islands which are developed around the main packet during the initial reshaping. 
These features can be so neatly resolved due to the strong coherence properties of polaritons.
Indeed such coherence is maintained during the interference phenomena between nonlinearly-induced counter-propagating flows.
In particular, the faster [$> v_g(k_x)$] leading and slower [$< v_g(k_x)$] trailing sub-packets 
represent the FWM states that are initially created at smaller ($k_x-\delta k_x$) and larger ($k_x+\delta k_x$) longitudinal momentum, respectively.
The counter-intuitive association between the group velocity and momentum differentials are due to the negative curvature of the dispersion.
Instead, the circular shape of the interference between the three packets suggests that the two excited FWM states 
have a larger transverse extension in momentum space ($k_y$) with respect to the main one.
In Fig.~\ref{dark.sol.}(a) we report both the density and phase longitudinal profiles (corresponding to $P=75~\mu\text{W}$ and $t = 7.5~\text{ps}$) to highlight the presence of two sharp $\pi$-jumps in front and behind the main packet,
in a perfect spatial correspondence to dark dips in the density profile.
The ignition time of such $\pi$-jump is also visible in the time space charts of Fig.~\ref{chart}(\textbf{c}) at $t\approx7.5-8.0~\text{ps}$,
that is also the time of vortex-antivortex pairs generation in real-space (see the main sequence in Fig.~\ref{opt.eff.}).
It is an interesting detail to note that the appearance of such dark-line itself 
looks as a couple of phase-singularities (quantum vortex) in the time-space domain of Fig.~\ref{chart}(\textbf{c}).
In general, the nonlinear self-development of a $\pi$-jump may be a signature of a dark soliton~\cite{dominici_vortex_2015,dominici_real-space_2015}, which can be sustained in two-dimensional condensates upon repulsive interactions~\cite{rodrigues_nodeless_2014}.
In Fig.~\ref{dark.sol.}(b) we report the evolution of the phase profiles at equidistant time frames (every $2.5~\text{ps}$). The profiles evidence how the sharp $\pi$-jump is only present in a given frame at early times, before smoothing down as expected due to the loss of intensity.  
Hence we may conclude that the dark soliton is just a transient structure here, formed due to the nonlinear way we ignite the XW in the polariton fluid and it is then washed out, without representing an intrinsic feature of the XW itself, as opposite to the longitudinal localization which is instead preserved in time.\\

\noindent \textbf{CONCLUSIONS}\\
\noindent
We have experimentally demonstrated the possibility to excite a peculiar class of traveling localized wavepackets, called X-waves, in two-dimensional exciton-polariton fluids.  
Self-generation of an X-wave out of a Gaussian excitation spot is obtained via a weakly nonlinear process asymmetric with respect to two directions of the non-parabolic polariton dispersion.
The dynamics of the packet are observed by means of ultrafast imaging, revealing a propagation over tens of micrometers, only limited by the polariton dissipation. 
We have tuned both the nonlinearity and the injected in-plane momentum to achieve the optimal effect and preserve the longitudinal localization even when density fades away. 
Different degrees of superluminality have been achieved and associated to the variable transverse angular aperture induced by the nonlinear process in its early stage.
Polariton based all-optical platforms are devised as robust candidates for the study of fundamental science connected to
two-dimensional X-wavepackets and possible future applications exploiting them in signal propagation.

Alternative 2D plaforms are represented by, e.g., multilayer stacks supporting Bloch surface waves (BSWs) at the external interface. 
These surface modes naturally exhibit very large in-plane speed---which converts into long range propagation---with negative mass dispersion 
and have shown exploitable nonlinearity upon coupling with an organic layer, stable up to room temperature~\cite{lerario_high-speed_2016}.
BSW are growing competitive systems compared to surface plasmon resonance for label-free high sensitivity biosensing.~\cite{sinibaldi_direct_2012}
Such systems also offer the possibility to easily pattern the external open surface to realize planar guiding or focusing elements~\cite{yu_manipulating_2014},
or even tilted top-grating launching diffraction-free surface waves~\cite{wang_diffraction-free_2017}, 
analogue to what has been previously realized with plasmonic systems~\cite{lin_cosine-gauss_2012}.
Hence, BSW polaritons are a natural evolution for the study of X-wave pulse propagation over hundreds of $\mu\text{m}$ distances and their exploitation
for novel 2D optical tweezers and sensing combined functionalities. 

Both QW-MC and BSW polariton
platforms represent nanophotonic technologies which are characterized 
by a strong and tunable $\chi^{(3)}$ nonlinearity resulting from polariton-polariton interaction. The third-order nonlinearity governs not only the four-wave mixing process but also other useful phenomena, such as self- and cross-Kerr modulation. Thus, we expect that they will be highly pronounced in our polariton superfluid. Tunable, efficient nonlinear interaction is a `holy grail' in photonic and optical systems~\cite{kishida_gigantic_2000,deng_four-wave_1999} and quantum computing~\cite{gottesman_heisenberg_1998}, in building optical gates necessary to construct a quantum computer.\\

\vspace{1mm}

\begin{small}


\noindent \textbf{AUTHORS’ CONTRIBUTIONS}\\
\noindent
L.D.~and M.M.~proposed the experiment.
A.G., L.D., D.B., M.D.G., G.G., and D.S.~set up the laboratory configuration.
A.G.~and L.D.~performed the experiments and analyzed the data. 
O.V., M.M., and M.S.~developed the theory, performed numerical simulations and provided the theoretical interpretations.
All the authors discussed the results. 
A.G., L.D., M.M., M.S., and D.S.~wrote the manuscript.  
D.S.~supervised the research.\\

\noindent \textbf{ACKNOWLEDGEMENTS}\\
\begin{small}
\noindent We thank R.~Houdr\'{e} and A.~Bramati for the growth and expertise on the microcavity device. 
AG, LD, DB, MDG, GG and DS are supported by the European Research Council POLAFLOW Grant 308136 and the Italian MIUR project Beyond Nano.
MS and OV are supported by the NCN grant No.~2012/04/M/ST2/00789 and MNiSW Iuventus Plus project No.~IP 2014 044873.
MS acknowledges support from the FNP project FIRST TEAM/2016-2/17.
MM acknowledges support from the NCN grant 2015/17/B/ST3/02273.\\
\end{small}

\noindent \textbf{Supplementary Material}\\
\noindent
\href{https://drive.google.com/open?id=0B0QCllnLqdyBUXdUeE0xX1JfYnc}
{Supplementary Movie S1}:
Experimental dynamics of the optimized X-wave generation and propagation as described in Fig.~2. The time step is $0.5~\text{ps}$.\\
\href{https://drive.google.com/open?id=0B0QCllnLqdyBUVQwVjU2VUxHUzg}
{Supplementary Movie S2}:
Experimental dynamics of the X-wave with a 3D rendering of the polariton amplitude.\\

\noindent \textbf{Cite as}
A.~Gianfrate, L.~Dominici, O.~Voronych, M.~Matuszewski, M.~Stobi\'{n}ska, D.~Ballarini, M.~ De Giorgi, G.~Gigli, D.~Sanvitto,
\emph{Superluminal X-waves in a polariton quantum fluid},
\href{https://www.nature.com/articles/lsa2017119}{​Light Sci.~Appl.~\textbf{7}, e17119 (2018)}.

\end{small}

\newpage

\def\bibsection{\section*{}} 
\noindent \textbf{Bibliography}


\begin{thebibliography}{61}%
\makeatletter
\providecommand \@ifxundefined [1]{%
 \@ifx{#1\undefined}
}%
\providecommand \@ifnum [1]{%
 \ifnum #1\expandafter \@firstoftwo
 \else \expandafter \@secondoftwo
 \fi
}%
\providecommand \@ifx [1]{%
 \ifx #1\expandafter \@firstoftwo
 \else \expandafter \@secondoftwo
 \fi
}%
\providecommand \natexlab [1]{#1}%
\providecommand \enquote  [1]{``#1''}%
\providecommand \bibnamefont  [1]{#1}%
\providecommand \bibfnamefont [1]{#1}%
\providecommand \citenamefont [1]{#1}%
\providecommand \href@noop [0]{\@secondoftwo}%
\providecommand \href [0]{\begingroup \@sanitize@url \@href}%
\providecommand \@href[1]{\@@startlink{#1}\@@href}%
\providecommand \@@href[1]{\endgroup#1\@@endlink}%
\providecommand \@sanitize@url [0]{\catcode `\\12\catcode `\$12\catcode
  `\&12\catcode `\#12\catcode `\^12\catcode `\_12\catcode `\%12\relax}%
\providecommand \@@startlink[1]{}%
\providecommand \@@endlink[0]{}%
\providecommand \url  [0]{\begingroup\@sanitize@url \@url }%
\providecommand \@url [1]{\endgroup\@href {#1}{\urlprefix }}%
\providecommand \urlprefix  [0]{URL }%
\providecommand \Eprint [0]{\href }%
\providecommand \doibase [0]{http://dx.doi.org/}%
\providecommand \selectlanguage [0]{\@gobble}%
\providecommand \bibinfo  [0]{\@secondoftwo}%
\providecommand \bibfield  [0]{\@secondoftwo}%
\providecommand \translation [1]{[#1]}%
\providecommand \BibitemOpen [0]{}%
\providecommand \bibitemStop [0]{}%
\providecommand \bibitemNoStop [0]{.\EOS\space}%
\providecommand \EOS [0]{\spacefactor3000\relax}%
\providecommand \BibitemShut  [1]{\csname bibitem#1\endcsname}%
\let\auto@bib@innerbib\@empty
\bibitem [{\citenamefont {Recami}\ \emph {et~al.}(2008)\citenamefont {Recami},
  \citenamefont {Zamboni-Rached},\ and\ \citenamefont
  {Hernandez-Figueroa}}]{hernandez-figueroa_localized_2008}%
  \BibitemOpen
  \bibfield  {author} {\bibinfo {author} {E. Recami}, \bibinfo {author} {M.
  Zamboni-Rached}, \bibinfo {author} {H.~E. Hernandez-Figueroa},\ }\bibinfo
  {title} {Localized {Waves}: {A} scientific and historical introduction},\ in\
  \href@noop {} {\emph {\bibinfo {booktitle} {Localized waves}}},\ \bibinfo
  {series and number} {Wiley series in microwave and optical engineering},\
  \bibinfo {editor} {edited by\ \bibinfo {editor} {H.~E.
  Hern\'{a}ndez-Figueroa}, \bibinfo {editor} {M. Zamboni-Rached}, \bibinfo
  {editor} {E. Recami}}\ (\bibinfo  {publisher} {Wiley-Interscience : IEEE
  Press},\ \bibinfo {address} {Hoboken, N.J},\ \bibinfo {year}
  {2008})\BibitemShut {NoStop}%
\bibitem [{\citenamefont {Recami}\ and\ \citenamefont
  {Zamboni-Rached}(2009)}]{recami_chapter_2009}%
  \BibitemOpen
  \bibfield  {author} {\bibinfo {author} {E. Recami}, \bibinfo {author} {M.
  Zamboni-Rached},\ }\bibinfo {title} {{Localized} {Waves}: {A} {Review}},\ in\
  \href {\doibase 10.1016/S1076-5670(08)01404-3} {\emph {\bibinfo {booktitle}
  {Advances in {Imaging} and {Electron} {Physics}}}},\ Vol.\ \bibinfo {volume}
  {156}\ (\bibinfo  {publisher} {Elsevier},\ \bibinfo {year} {2009})\
  Chap.~\bibinfo {chapter} {4}, pp.\ \bibinfo {pages} {235--353}\BibitemShut
  {NoStop}%
\bibitem [{\citenamefont {S\~{o}najalg}\ \emph {et~al.}(1997)\citenamefont
  {S\~{o}najalg}, \citenamefont {R\"{a}tsep},\ and\ \citenamefont
  {Saari}}]{sonajalg_demonstration_1997}%
  \BibitemOpen
  \bibfield  {author} {\bibinfo {author} {H. S\~{o}najalg}, \bibinfo {author}
  {M. R\"{a}tsep}, \bibinfo {author} {P. Saari},\ }\bibfield  {title} {\emph
  {\bibinfo {title} {Demonstration of the {Bessel}-{X} pulse propagating with
  strong lateral and longitudinal localization in a dispersive medium},\
  }}\href {\doibase 10.1364/OL.22.000310} {\bibfield  {journal} {\bibinfo
  {journal} {Opt. Lett.}\ }\textbf {\bibinfo {volume} {22}},\ \bibinfo {pages}
  {310} (\bibinfo {year} {1997})}\BibitemShut {NoStop}%
\bibitem [{\citenamefont {Durnin}\ \emph {et~al.}(1987)\citenamefont {Durnin},
  \citenamefont {Miceli},\ and\ \citenamefont
  {Eberly}}]{durnin_diffraction-free_1987}%
  \BibitemOpen
  \bibfield  {author} {\bibinfo {author} {J. Durnin}, \bibinfo {author} {J.~J.
  Miceli}, \bibinfo {author} {J.~H. Eberly},\ }\bibfield  {title} {\emph
  {\bibinfo {title} {Diffraction-free beams},\ }}\href {\doibase
  10.1103/PhysRevLett.58.1499} {\bibfield  {journal} {\bibinfo  {journal}
  {Phys. Rev. Lett.}\ }\textbf {\bibinfo {volume} {58}},\ \bibinfo {pages}
  {1499} (\bibinfo {year} {1987})}\BibitemShut {NoStop}%
\bibitem [{\citenamefont {Salo}\ \emph {et~al.}(2000)\citenamefont {Salo},
  \citenamefont {Fagerholm}, \citenamefont {Friberg},\ and\ \citenamefont
  {Salomaa}}]{salo_unified_2000}%
  \BibitemOpen
  \bibfield  {author} {\bibinfo {author} {J. Salo}, \bibinfo {author} {J.
  Fagerholm}, \bibinfo {author} {A.~T. Friberg}, \bibinfo {author} {M.~M.
  Salomaa},\ }\bibfield  {title} {\emph {\bibinfo {title} {Unified description
  of nondiffracting {X} and {Y} waves},\ }}\href {\doibase
  10.1103/PhysRevE.62.4261} {\bibfield  {journal} {\bibinfo  {journal} {Phys.
  Rev. E}\ }\textbf {\bibinfo {volume} {62}},\ \bibinfo {pages} {4261}
  (\bibinfo {year} {2000})}\BibitemShut {NoStop}%
\bibitem [{\citenamefont {Lu}\ and\ \citenamefont
  {Greenleaf}(1990)}]{lu_ultrasonic_1990}%
  \BibitemOpen
  \bibfield  {author} {\bibinfo {author} {J.~Y. Lu}, \bibinfo {author} {J.~F.
  Greenleaf},\ }\bibfield  {title} {\emph {\bibinfo {title} {Ultrasonic
  nondiffracting transducer for medical imaging},\ }}\href {\doibase
  10.1109/58.105250} {\bibfield  {journal} {\bibinfo  {journal} {IEEE Trans.
  Ultrason. Ferroelect. Freq. Control}\ }\textbf {\bibinfo {volume} {37}},\
  \bibinfo {pages} {438} (\bibinfo {year} {1990})}\BibitemShut {NoStop}%
\bibitem [{\citenamefont {Lu}\ and\ \citenamefont
  {Greenleaf}(1992)}]{lu_nondiffracting_1992}%
  \BibitemOpen
  \bibfield  {author} {\bibinfo {author} {J.~Y. Lu}, \bibinfo {author} {J.~F.
  Greenleaf},\ }\bibfield  {title} {\emph {\bibinfo {title} {Nondiffracting {X}
  waves-exact solutions to free-space scalar wave equation and their finite
  aperture realizations},\ }}\href {\doibase 10.1109/58.166806} {\bibfield
  {journal} {\bibinfo  {journal} {IEEE Trans. Ultrason. Ferroelect. Freq.
  Control}\ }\textbf {\bibinfo {volume} {39}},\ \bibinfo {pages} {19} (\bibinfo
  {year} {1992})}\BibitemShut {NoStop}%
\bibitem [{\citenamefont {Yalizay}\ \emph {et~al.}(2012)\citenamefont
  {Yalizay}, \citenamefont {Ersoy}, \citenamefont {Soylu},\ and\ \citenamefont
  {Akturk}}]{yalizay_fabrication_2012}%
  \BibitemOpen
  \bibfield  {author} {\bibinfo {author} {B. Yalizay}, \bibinfo {author} {T.
  Ersoy}, \bibinfo {author} {B. Soylu}, \bibinfo {author} {S. Akturk},\
  }\bibfield  {title} {\emph {\bibinfo {title} {Fabrication of nanometer-size
  structures in metal thin films using femtosecond laser {Bessel} beams},\
  }}\href {\doibase 10.1063/1.3678030} {\bibfield  {journal} {\bibinfo
  {journal} {Appl. Phys. Lett.}\ }\textbf {\bibinfo {volume} {100}},\ \bibinfo
  {pages} {031104} (\bibinfo {year} {2012})}\BibitemShut {NoStop}%
\bibitem [{\citenamefont {Zamboni-Rached}\ \emph {et~al.}(2002)\citenamefont
  {Zamboni-Rached}, \citenamefont {Recami},\ and\ \citenamefont
  {Hern\'{a}ndez-Figueroa}}]{zamboni-rached_new_2002}%
  \BibitemOpen
  \bibfield  {author} {\bibinfo {author} {M. Zamboni-Rached}, \bibinfo {author}
  {E. Recami}, \bibinfo {author} {H. Hern\'{a}ndez-Figueroa},\ }\bibfield
  {title} {\emph {\bibinfo {title} {New localized {Superluminal} solutions to
  the wave equations with finite total energies and arbitrary frequencies},\
  }}\href {\doibase 10.1140/epjd/e2002-00198-7} {\bibfield  {journal} {\bibinfo
   {journal} {Eur. Phys. J. D}\ }\textbf {\bibinfo {volume} {21}},\ \bibinfo
  {pages} {217} (\bibinfo {year} {2002})}\BibitemShut {NoStop}%
\bibitem [{\citenamefont {Zamboni-Rached}\ \emph {et~al.}(2003)\citenamefont
  {Zamboni-Rached}, \citenamefont {Fontana},\ and\ \citenamefont
  {Recami}}]{zamboni-rached_superluminal_2003}%
  \BibitemOpen
  \bibfield  {author} {\bibinfo {author} {M. Zamboni-Rached}, \bibinfo {author}
  {F. Fontana}, \bibinfo {author} {E. Recami},\ }\bibfield  {title} {\emph
  {\bibinfo {title} {Superluminal localized solutions to {Maxwell} equations
  propagating along a waveguide: {The} finite-energy case},\ }}\href {\doibase
  10.1103/PhysRevE.67.036620} {\bibfield  {journal} {\bibinfo  {journal} {Phys.
  Rev. E}\ }\textbf {\bibinfo {volume} {67}},\ \bibinfo {pages} {036620}
  (\bibinfo {year} {2003})}\BibitemShut {NoStop}%
\bibitem [{\citenamefont {Conti}\ and\ \citenamefont
  {Trillo}(2004)}]{conti_nonspreading_2004}%
  \BibitemOpen
  \bibfield  {author} {\bibinfo {author} {C. Conti}, \bibinfo {author} {S.
  Trillo},\ }\bibfield  {title} {\emph {\bibinfo {title} {Nonspreading {Wave}
  {Packets} in {Three} {Dimensions} {Formed} by an {Ultracold} {Bose} {Gas} in
  an {Optical} {Lattice}},\ }}\href {\doibase 10.1103/PhysRevLett.92.120404}
  {\bibfield  {journal} {\bibinfo  {journal} {Phys. Rev. Lett.}\ }\textbf
  {\bibinfo {volume} {92}},\ \bibinfo {pages} {120404} (\bibinfo {year}
  {2004})}\BibitemShut {NoStop}%
\bibitem [{\citenamefont {Voronych}\ \emph {et~al.}(2016)\citenamefont
  {Voronych}, \citenamefont {Buraczewski}, \citenamefont {Matuszewski},\ and\
  \citenamefont {Stobi\'{n}ska}}]{voronych_exciton-polariton_2016}%
  \BibitemOpen
  \bibfield  {author} {\bibinfo {author} {O. Voronych}, \bibinfo {author} {A.
  Buraczewski}, \bibinfo {author} {M. Matuszewski}, \bibinfo {author} {M.
  Stobi\'{n}ska},\ }\bibfield  {title} {\emph {\bibinfo {title}
  {Exciton-polariton localized wave packets in a microcavity},\ }}\href
  {\doibase 10.1103/PhysRevB.93.245310} {\bibfield  {journal} {\bibinfo
  {journal} {Phys. Rev. B}\ }\textbf {\bibinfo {volume} {93}},\ \bibinfo
  {pages} {245310} (\bibinfo {year} {2016})}\BibitemShut {NoStop}%
\bibitem [{\citenamefont {Efremidis}\ \emph {et~al.}(2009)\citenamefont
  {Efremidis}, \citenamefont {Siviloglou},\ and\ \citenamefont
  {Christodoulides}}]{efremidis_exact_2009}%
  \BibitemOpen
  \bibfield  {author} {\bibinfo {author} {N.~K. Efremidis}, \bibinfo {author}
  {G.~A. Siviloglou}, \bibinfo {author} {D.~N. Christodoulides},\ }\bibfield
  {title} {\emph {\bibinfo {title} {Exact {X}-wave solutions of the hyperbolic
  nonlinear {Schr\"{o}dinger} equation with a supporting potential},\ }}\href
  {\doibase 10.1016/j.physleta.2009.09.008} {\bibfield  {journal} {\bibinfo
  {journal} {Phys. Lett. A}\ }\textbf {\bibinfo {volume} {373}},\ \bibinfo
  {pages} {4073} (\bibinfo {year} {2009})}\BibitemShut {NoStop}%
\bibitem [{\citenamefont {Conti}(2004)}]{conti_generation_2004}%
  \BibitemOpen
  \bibfield  {author} {\bibinfo {author} {C. Conti},\ }\bibfield  {title}
  {\emph {\bibinfo {title} {Generation and nonlinear dynamics of {X} waves of
  the {Schr\"{o}dinger} equation},\ }}\href {\doibase
  10.1103/PhysRevE.70.046613} {\bibfield  {journal} {\bibinfo  {journal} {Phys.
  Rev. E}\ }\textbf {\bibinfo {volume} {70}},\ \bibinfo {pages} {046613}
  (\bibinfo {year} {2004})}\BibitemShut {NoStop}%
\bibitem [{\citenamefont {Sedov}\ \emph {et~al.}(2015)\citenamefont {Sedov},
  \citenamefont {Iorsh}, \citenamefont {Arakelian}, \citenamefont {Alodjants},\
  and\ \citenamefont {Kavokin}}]{sedov_hyperbolic_2015}%
  \BibitemOpen
  \bibfield  {author} {\bibinfo {author} {E.~S. Sedov}, \bibinfo {author} {I.
  Iorsh}, \bibinfo {author} {S. Arakelian}, \bibinfo {author} {A. Alodjants},
  \bibinfo {author} {A. Kavokin},\ }\bibfield  {title} {\emph {\bibinfo {title}
  {Hyperbolic {Metamaterials} with {Bragg} {Polaritons}},\ }}\href {\doibase
  10.1103/PhysRevLett.114.237402} {\bibfield  {journal} {\bibinfo  {journal}
  {Physical Review Letters}\ }\textbf {\bibinfo {volume} {114}} (\bibinfo
  {year} {2015}),\ 10.1103/PhysRevLett.114.237402}\BibitemShut {NoStop}%
\bibitem [{\citenamefont {Couairon}\ \emph {et~al.}(2006)\citenamefont
  {Couairon}, \citenamefont {Gai\v{z}auskas}, \citenamefont {Faccio},
  \citenamefont {Dubietis},\ and\ \citenamefont
  {Di~Trapani}}]{couairon_nonlinear_2006}%
  \BibitemOpen
  \bibfield  {author} {\bibinfo {author} {A. Couairon}, \bibinfo {author} {E.
  Gai\v{z}auskas}, \bibinfo {author} {D. Faccio}, \bibinfo {author} {A.
  Dubietis}, \bibinfo {author} {P. Di~Trapani},\ }\bibfield  {title} {\emph
  {\bibinfo {title} {Nonlinear {X}-wave formation by femtosecond filamentation
  in {Kerr} media},\ }}\href {\doibase 10.1103/PhysRevE.73.016608} {\bibfield
  {journal} {\bibinfo  {journal} {Phys. Rev. E}\ }\textbf {\bibinfo {volume}
  {73}},\ \bibinfo {pages} {016608} (\bibinfo {year} {2006})}\BibitemShut
  {NoStop}%
\bibitem [{\citenamefont {Kolesik}\ \emph {et~al.}(2004)\citenamefont
  {Kolesik}, \citenamefont {Wright},\ and\ \citenamefont
  {Moloney}}]{kolesik_dynamic_2004}%
  \BibitemOpen
  \bibfield  {author} {\bibinfo {author} {M. Kolesik}, \bibinfo {author} {E.~M.
  Wright}, \bibinfo {author} {J.~V. Moloney},\ }\bibfield  {title} {\emph
  {\bibinfo {title} {Dynamic {Nonlinear} {X} {Waves} for {Femtosecond} {Pulse}
  {Propagation} in {Water}},\ }}\href {\doibase 10.1103/PhysRevLett.92.253901}
  {\bibfield  {journal} {\bibinfo  {journal} {Phys. Rev. Lett.}\ }\textbf
  {\bibinfo {volume} {92}},\ \bibinfo {pages} {253901} (\bibinfo {year}
  {2004})}\BibitemShut {NoStop}%
\bibitem [{\citenamefont {Conti}\ \emph {et~al.}(2003)\citenamefont {Conti},
  \citenamefont {Trillo}, \citenamefont {Di~Trapani}, \citenamefont {Valiulis},
  \citenamefont {Piskarskas}, \citenamefont {Jedrkiewicz},\ and\ \citenamefont
  {Trull}}]{conti_nonlinear_2003}%
  \BibitemOpen
  \bibfield  {author} {\bibinfo {author} {C. Conti}, \bibinfo {author} {S.
  Trillo}, \bibinfo {author} {P. Di~Trapani}, \bibinfo {author} {G. Valiulis},
  \bibinfo {author} {A. Piskarskas}, \bibinfo {author} {O. Jedrkiewicz},
  \bibinfo {author} {J. Trull},\ }\bibfield  {title} {\emph {\bibinfo {title}
  {Nonlinear {Electromagnetic} {X} {Waves}},\ }}\href {\doibase
  10.1103/PhysRevLett.90.170406} {\bibfield  {journal} {\bibinfo  {journal}
  {Phys. Rev. Lett.}\ }\textbf {\bibinfo {volume} {90}},\ \bibinfo {pages}
  {170406} (\bibinfo {year} {2003})}\BibitemShut {NoStop}%
\bibitem [{\citenamefont {Di~Trapani}\ \emph {et~al.}(2003)\citenamefont
  {Di~Trapani}, \citenamefont {Valiulis}, \citenamefont {Piskarskas},
  \citenamefont {Jedrkiewicz}, \citenamefont {Trull}, \citenamefont {Conti},\
  and\ \citenamefont {Trillo}}]{di_trapani_spontaneously_2003}%
  \BibitemOpen
  \bibfield  {author} {\bibinfo {author} {P. Di~Trapani}, \bibinfo {author} {G.
  Valiulis}, \bibinfo {author} {A. Piskarskas}, \bibinfo {author} {O.
  Jedrkiewicz}, \bibinfo {author} {J. Trull}, \bibinfo {author} {C. Conti},
  \bibinfo {author} {S. Trillo},\ }\bibfield  {title} {\emph {\bibinfo {title}
  {Spontaneously {Generated} {X}-{Shaped} {Light} {Bullets}},\ }}\href
  {\doibase 10.1103/PhysRevLett.91.093904} {\bibfield  {journal} {\bibinfo
  {journal} {Phys. Rev. Lett.}\ }\textbf {\bibinfo {volume} {91}},\ \bibinfo
  {pages} {093904} (\bibinfo {year} {2003})}\BibitemShut {NoStop}%
\bibitem [{\citenamefont {Ciattoni}\ and\ \citenamefont
  {Conti}(2007)}]{ciattoni_quantum_2007}%
  \BibitemOpen
  \bibfield  {author} {\bibinfo {author} {A. Ciattoni}, \bibinfo {author} {C.
  Conti},\ }\bibfield  {title} {\emph {\bibinfo {title} {Quantum
  electromagnetic {X} waves},\ }}\href {\doibase 10.1364/JOSAB.24.002195}
  {\bibfield  {journal} {\bibinfo  {journal} {J. Opt. Soc. Am. B}\ }\textbf
  {\bibinfo {volume} {24}},\ \bibinfo {pages} {2195} (\bibinfo {year}
  {2007})}\BibitemShut {NoStop}%
\bibitem [{\citenamefont {Sanvitto}\ and\ \citenamefont
  {K\'{e}na-Cohen}(2016)}]{sanvitto_road_2016}%
  \BibitemOpen
  \bibfield  {author} {\bibinfo {author} {D. Sanvitto}, \bibinfo {author} {S.
  K\'{e}na-Cohen},\ }\bibfield  {title} {\emph {\bibinfo {title} {The road
  towards polaritonic devices},\ }}\href {\doibase 10.1038/nmat4668} {\bibfield
   {journal} {\bibinfo  {journal} {Nat. Mater.}\ }\textbf {\bibinfo {volume}
  {15}},\ \bibinfo {pages} {1061} (\bibinfo {year} {2016})}\BibitemShut
  {NoStop}%
\bibitem [{\citenamefont {Byrnes}\ \emph {et~al.}(2014)\citenamefont {Byrnes},
  \citenamefont {Kim},\ and\ \citenamefont
  {Yamamoto}}]{byrnes_excitonpolariton_2014}%
  \BibitemOpen
  \bibfield  {author} {\bibinfo {author} {T. Byrnes}, \bibinfo {author} {N.~Y.
  Kim}, \bibinfo {author} {Y. Yamamoto},\ }\bibfield  {title} {\emph {\bibinfo
  {title} {Exciton--polariton condensates},\ }}\href {\doibase
  10.1038/nphys3143} {\bibfield  {journal} {\bibinfo  {journal} {Nat. Phys.}\
  }\textbf {\bibinfo {volume} {10}},\ \bibinfo {pages} {803} (\bibinfo {year}
  {2014})}\BibitemShut {NoStop}%
\bibitem [{\citenamefont {Dagvadorj}\ \emph {et~al.}(2015)\citenamefont
  {Dagvadorj}, \citenamefont {Fellows}, \citenamefont {Matyja\'{s}kiewicz},
  \citenamefont {Marchetti}, \citenamefont {Carusotto},\ and\ \citenamefont
  {Szyma\'{n}ska}}]{dagvadorj_nonequilibrium_2015}%
  \BibitemOpen
  \bibfield  {author} {\bibinfo {author} {G. Dagvadorj}, \bibinfo {author}
  {J.~M. Fellows}, \bibinfo {author} {S. Matyja\'{s}kiewicz}, \bibinfo {author}
  {F.~M. Marchetti}, \bibinfo {author} {I. Carusotto}, \bibinfo {author} {M.~H.
  Szyma\'{n}ska},\ }\bibfield  {title} {\emph {\bibinfo {title} {Nonequilibrium
  {Phase} {Transition} in a {Two}-{Dimensional} {Driven} {Open} {Quantum}
  {System}},\ }}\href {\doibase 10.1103/PhysRevX.5.041028} {\bibfield
  {journal} {\bibinfo  {journal} {Phys. Rev. X}\ }\textbf {\bibinfo {volume}
  {5}},\ \bibinfo {pages} {041028} (\bibinfo {year} {2015})}\BibitemShut
  {NoStop}%
\bibitem [{\citenamefont {Deng}\ \emph {et~al.}(2010)\citenamefont {Deng},
  \citenamefont {Haug},\ and\ \citenamefont
  {Yamamoto}}]{deng_exciton-polariton_2010}%
  \BibitemOpen
  \bibfield  {author} {\bibinfo {author} {H. Deng}, \bibinfo {author} {H.
  Haug}, \bibinfo {author} {Y. Yamamoto},\ }\bibfield  {title} {\emph {\bibinfo
  {title} {Exciton-polariton {Bose}-{Einstein} condensation},\ }}\href
  {\doibase 10.1103/RevModPhys.82.1489} {\bibfield  {journal} {\bibinfo
  {journal} {Rev. Mod. Phys.}\ }\textbf {\bibinfo {volume} {82}},\ \bibinfo
  {pages} {1489} (\bibinfo {year} {2010})}\BibitemShut {NoStop}%
\bibitem [{\citenamefont {Amo}\ \emph {et~al.}(2009{\natexlab{a}})\citenamefont
  {Amo}, \citenamefont {Sanvitto}, \citenamefont {Laussy}, \citenamefont
  {Ballarini}, \citenamefont {Valle}, \citenamefont {Martin}, \citenamefont
  {Lema\^{i}tre}, \citenamefont {Bloch}, \citenamefont {Krizhanovskii},
  \citenamefont {Skolnick}, \citenamefont {Tejedor},\ and\ \citenamefont
  {Vi\~{n}a}}]{amo_collective_2009}%
  \BibitemOpen
  \bibfield  {author} {\bibinfo {author} {A. Amo}, \bibinfo {author} {D.
  Sanvitto}, \bibinfo {author} {F.~P. Laussy}, \bibinfo {author} {D.
  Ballarini}, \bibinfo {author} {E.~d. Valle}, \bibinfo {author} {M.~D.
  Martin}, \bibinfo {author} {A. Lema\^{i}tre}, \bibinfo {author} {J. Bloch},
  \bibinfo {author} {D.~N. Krizhanovskii}, \bibinfo {author} {M.~S. Skolnick},
  \bibinfo {author} {C. Tejedor}, \bibinfo {author} {L. Vi\~{n}a},\ }\bibfield
  {title} {\emph {\bibinfo {title} {Collective fluid dynamics of a polariton
  condensate in a semiconductor microcavity},\ }}\href {\doibase
  10.1038/nature07640} {\bibfield  {journal} {\bibinfo  {journal} {Nature}\
  }\textbf {\bibinfo {volume} {457}},\ \bibinfo {pages} {291} (\bibinfo {year}
  {2009}{\natexlab{a}})}\BibitemShut {NoStop}%
\bibitem [{\citenamefont {Kasprzak}\ \emph {et~al.}(2006)\citenamefont
  {Kasprzak}, \citenamefont {Richard}, \citenamefont {Kundermann},
  \citenamefont {Baas}, \citenamefont {Jeambrun}, \citenamefont {Keeling},
  \citenamefont {Marchetti}, \citenamefont {Szyma\'{n}ska}, \citenamefont
  {Andr\'{e}}, \citenamefont {Staehli}, \citenamefont {Savona}, \citenamefont
  {Littlewood}, \citenamefont {Deveaud},\ and\ \citenamefont
  {Dang}}]{kasprzak_boseeinstein_2006}%
  \BibitemOpen
  \bibfield  {author} {\bibinfo {author} {J. Kasprzak}, \bibinfo {author} {M.
  Richard}, \bibinfo {author} {S. Kundermann}, \bibinfo {author} {A. Baas},
  \bibinfo {author} {P. Jeambrun}, \bibinfo {author} {J.~M.~J. Keeling},
  \bibinfo {author} {F.~M. Marchetti}, \bibinfo {author} {M.~H. Szyma\'{n}ska},
  \bibinfo {author} {R. Andr\'{e}}, \bibinfo {author} {J.~L. Staehli}, \bibinfo
  {author} {V. Savona}, \bibinfo {author} {P.~B. Littlewood}, \bibinfo {author}
  {B. Deveaud}, \bibinfo {author} {L.~S. Dang},\ }\bibfield  {title} {\emph
  {\bibinfo {title} {Bose-{Einstein} condensation of exciton polaritons},\
  }}\href {\doibase 10.1038/nature05131} {\bibfield  {journal} {\bibinfo
  {journal} {Nature}\ }\textbf {\bibinfo {volume} {443}},\ \bibinfo {pages}
  {409} (\bibinfo {year} {2006})}\BibitemShut {NoStop}%
\bibitem [{\citenamefont {Balili}\ \emph {et~al.}(2007)\citenamefont {Balili},
  \citenamefont {Hartwell}, \citenamefont {Snoke}, \citenamefont {Pfeiffer},\
  and\ \citenamefont {West}}]{balili_bose-einstein_2007}%
  \BibitemOpen
  \bibfield  {author} {\bibinfo {author} {R. Balili}, \bibinfo {author} {V.
  Hartwell}, \bibinfo {author} {D. Snoke}, \bibinfo {author} {L. Pfeiffer},
  \bibinfo {author} {K. West},\ }\bibfield  {title} {\emph {\bibinfo {title}
  {Bose-{Einstein} {Condensation} of {Microcavity} {Polaritons} in a {Trap}},\
  }}\href {\doibase 10.1126/science.1140990} {\bibfield  {journal} {\bibinfo
  {journal} {Science}\ }\textbf {\bibinfo {volume} {316}},\ \bibinfo {pages}
  {1007} (\bibinfo {year} {2007})}\BibitemShut {NoStop}%
\bibitem [{\citenamefont {Kavokin}\ \emph {et~al.}(2017)\citenamefont
  {Kavokin}, \citenamefont {Baumberg}, \citenamefont {Malpuech},\ and\
  \citenamefont {Laussy}}]{kavokin_microcavities_2017}%
  \BibitemOpen
  \bibfield  {author} {\bibinfo {author} {A.~V. Kavokin}, \bibinfo {author}
  {J.~J. Baumberg}, \bibinfo {author} {G. Malpuech}, \bibinfo {author} {F.~P.
  Laussy},\ }\href@noop {} {\emph {\bibinfo {title} {Microcavities}}},\
  \bibinfo {edition} {2nd}\ ed.,\ Series on {Semiconductor} {Science} and
  {Technology}\ (\bibinfo  {publisher} {Oxford University Press},\ \bibinfo
  {address} {Oxford, New York},\ \bibinfo {year} {2017})\BibitemShut {NoStop}%
\bibitem [{\citenamefont {Colas}\ and\ \citenamefont
  {Laussy}(2016)}]{colas_self-interfering_2016}%
  \BibitemOpen
  \bibfield  {author} {\bibinfo {author} {D. Colas}, \bibinfo {author} {F.~P.
  Laussy},\ }\bibfield  {title} {\emph {\bibinfo {title} {Self-{Interfering}
  {Wave} {Packets}},\ }}\href {\doibase 10.1103/PhysRevLett.116.026401}
  {\bibfield  {journal} {\bibinfo  {journal} {Phys. Rev. Lett.}\ }\textbf
  {\bibinfo {volume} {116}},\ \bibinfo {pages} {026401} (\bibinfo {year}
  {2016})}\BibitemShut {NoStop}%
\bibitem [{\citenamefont {Walker}\ \emph {et~al.}(2015)\citenamefont {Walker},
  \citenamefont {Tinkler}, \citenamefont {Skryabin}, \citenamefont {Yulin},
  \citenamefont {Royall}, \citenamefont {Farrer}, \citenamefont {Ritchie},
  \citenamefont {Skolnick},\ and\ \citenamefont
  {Krizhanovskii}}]{walker_ultra-low-power_2015}%
  \BibitemOpen
  \bibfield  {author} {\bibinfo {author} {P.~M. Walker}, \bibinfo {author} {L.
  Tinkler}, \bibinfo {author} {D.~V. Skryabin}, \bibinfo {author} {A. Yulin},
  \bibinfo {author} {B. Royall}, \bibinfo {author} {I. Farrer}, \bibinfo
  {author} {D.~A. Ritchie}, \bibinfo {author} {M.~S. Skolnick}, \bibinfo
  {author} {D.~N. Krizhanovskii},\ }\bibfield  {title} {\emph {\bibinfo {title}
  {Ultra-low-power hybrid light-matter solitons},\ }}\href {\doibase
  10.1038/ncomms9317} {\bibfield  {journal} {\bibinfo  {journal} {Nat.
  Commun.}\ }\textbf {\bibinfo {volume} {6}},\ \bibinfo {pages} {8317}
  (\bibinfo {year} {2015})}\BibitemShut {NoStop}%
\bibitem [{\citenamefont {Vladimirova}\ \emph {et~al.}(2010)\citenamefont
  {Vladimirova}, \citenamefont {Cronenberger}, \citenamefont {Scalbert},
  \citenamefont {Kavokin}, \citenamefont {Miard}, \citenamefont {Lema\^{i}tre},
  \citenamefont {Bloch}, \citenamefont {Solnyshkov}, \citenamefont {Malpuech},\
  and\ \citenamefont {Kavokin}}]{vladimirova_polariton-polariton_2010}%
  \BibitemOpen
  \bibfield  {author} {\bibinfo {author} {M. Vladimirova}, \bibinfo {author}
  {S. Cronenberger}, \bibinfo {author} {D. Scalbert}, \bibinfo {author} {K.~V.
  Kavokin}, \bibinfo {author} {A. Miard}, \bibinfo {author} {A. Lema\^{i}tre},
  \bibinfo {author} {J. Bloch}, \bibinfo {author} {D. Solnyshkov}, \bibinfo
  {author} {G. Malpuech}, \bibinfo {author} {A.~V. Kavokin},\ }\bibfield
  {title} {\emph {\bibinfo {title} {Polariton-polariton interaction constants
  in microcavities},\ }}\href {\doibase 10.1103/PhysRevB.82.075301} {\bibfield
  {journal} {\bibinfo  {journal} {Phys. Rev. B}\ }\textbf {\bibinfo {volume}
  {82}},\ \bibinfo {pages} {075301} (\bibinfo {year} {2010})}\BibitemShut
  {NoStop}%
\bibitem [{\citenamefont {Amo}\ \emph {et~al.}(2009{\natexlab{b}})\citenamefont
  {Amo}, \citenamefont {Lefr\`{e}re}, \citenamefont {Pigeon}, \citenamefont
  {Adrados}, \citenamefont {Ciuti}, \citenamefont {Carusotto}, \citenamefont
  {Houdr\'{e}}, \citenamefont {Giacobino},\ and\ \citenamefont
  {Bramati}}]{amo_superfluidity_2009}%
  \BibitemOpen
  \bibfield  {author} {\bibinfo {author} {A. Amo}, \bibinfo {author} {J.
  Lefr\`{e}re}, \bibinfo {author} {S. Pigeon}, \bibinfo {author} {C. Adrados},
  \bibinfo {author} {C. Ciuti}, \bibinfo {author} {I. Carusotto}, \bibinfo
  {author} {R. Houdr\'{e}}, \bibinfo {author} {E. Giacobino}, \bibinfo {author}
  {A. Bramati},\ }\bibfield  {title} {\emph {\bibinfo {title} {Superfluidity of
  polaritons in semiconductor microcavities},\ }}\href {\doibase
  10.1038/nphys1364} {\bibfield  {journal} {\bibinfo  {journal} {Nat. Phys.}\
  }\textbf {\bibinfo {volume} {5}},\ \bibinfo {pages} {805} (\bibinfo {year}
  {2009}{\natexlab{b}})}\BibitemShut {NoStop}%
\bibitem [{\citenamefont {Berceanu}\ \emph {et~al.}(2015)\citenamefont
  {Berceanu}, \citenamefont {Dominici}, \citenamefont {Carusotto},
  \citenamefont {Ballarini}, \citenamefont {Cancellieri}, \citenamefont
  {Gigli}, \citenamefont {Szyma\'{n}ska}, \citenamefont {Sanvitto},\ and\
  \citenamefont {Marchetti}}]{berceanu_multicomponent_2015}%
  \BibitemOpen
  \bibfield  {author} {\bibinfo {author} {A.~C. Berceanu}, \bibinfo {author}
  {L. Dominici}, \bibinfo {author} {I. Carusotto}, \bibinfo {author} {D.
  Ballarini}, \bibinfo {author} {E. Cancellieri}, \bibinfo {author} {G. Gigli},
  \bibinfo {author} {M.~H. Szyma\'{n}ska}, \bibinfo {author} {D. Sanvitto},
  \bibinfo {author} {F.~M. Marchetti},\ }\bibfield  {title} {\emph {\bibinfo
  {title} {Multicomponent polariton superfluidity in the optical parametric
  oscillator regime},\ }}\href {\doibase 10.1103/PhysRevB.92.035307} {\bibfield
   {journal} {\bibinfo  {journal} {Phys. Rev. B}\ }\textbf {\bibinfo {volume}
  {92}},\ \bibinfo {pages} {035307} (\bibinfo {year} {2015})}\BibitemShut
  {NoStop}%
\bibitem [{\citenamefont {Amo}\ \emph {et~al.}(2011)\citenamefont {Amo},
  \citenamefont {Pigeon}, \citenamefont {Sanvitto}, \citenamefont {Sala},
  \citenamefont {Hivet}, \citenamefont {Carusotto}, \citenamefont {Pisanello},
  \citenamefont {Lemenager}, \citenamefont {Houdr\'{e}}, \citenamefont
  {Giacobino}, \citenamefont {Ciuti},\ and\ \citenamefont
  {Bramati}}]{amo_polariton_2011}%
  \BibitemOpen
  \bibfield  {author} {\bibinfo {author} {A. Amo}, \bibinfo {author} {S.
  Pigeon}, \bibinfo {author} {D. Sanvitto}, \bibinfo {author} {V.~G. Sala},
  \bibinfo {author} {R. Hivet}, \bibinfo {author} {I. Carusotto}, \bibinfo
  {author} {F. Pisanello}, \bibinfo {author} {G. Lemenager}, \bibinfo {author}
  {R. Houdr\'{e}}, \bibinfo {author} {E. Giacobino}, \bibinfo {author} {C.
  Ciuti}, \bibinfo {author} {A. Bramati},\ }\bibfield  {title} {\emph {\bibinfo
  {title} {Polariton {Superfluids} {Reveal} {Quantum} {Hydrodynamic}
  {Solitons}},\ }}\href {\doibase 10.1126/science.1202307} {\bibfield
  {journal} {\bibinfo  {journal} {Science}\ }\textbf {\bibinfo {volume}
  {332}},\ \bibinfo {pages} {1167} (\bibinfo {year} {2011})}\BibitemShut
  {NoStop}%
\bibitem [{\citenamefont {Sanvitto}\ \emph {et~al.}(2010)\citenamefont
  {Sanvitto}, \citenamefont {Marchetti}, \citenamefont {Szyma\'{n}ska},
  \citenamefont {Tosi}, \citenamefont {Baudisch}, \citenamefont {Laussy},
  \citenamefont {Krizhanovskii}, \citenamefont {Skolnick}, \citenamefont
  {Marrucci}, \citenamefont {Lema\^{i}tre}, \citenamefont {Bloch},
  \citenamefont {Tejedor},\ and\ \citenamefont
  {Vi\~{n}a}}]{sanvitto_persistent_2010}%
  \BibitemOpen
  \bibfield  {author} {\bibinfo {author} {D. Sanvitto}, \bibinfo {author}
  {F.~M. Marchetti}, \bibinfo {author} {M.~H. Szyma\'{n}ska}, \bibinfo {author}
  {G. Tosi}, \bibinfo {author} {M. Baudisch}, \bibinfo {author} {F.~P. Laussy},
  \bibinfo {author} {D.~N. Krizhanovskii}, \bibinfo {author} {M.~S. Skolnick},
  \bibinfo {author} {L. Marrucci}, \bibinfo {author} {A. Lema\^{i}tre},
  \bibinfo {author} {J. Bloch}, \bibinfo {author} {C. Tejedor}, \bibinfo
  {author} {L. Vi\~{n}a},\ }\bibfield  {title} {\emph {\bibinfo {title}
  {Persistent currents and quantized vortices in a polariton superfluid},\
  }}\href {\doibase 10.1038/nphys1668} {\bibfield  {journal} {\bibinfo
  {journal} {Nat. Phys.}\ }\textbf {\bibinfo {volume} {6}},\ \bibinfo {pages}
  {527} (\bibinfo {year} {2010})}\BibitemShut {NoStop}%
\bibitem [{\citenamefont {Whittaker}\ \emph {et~al.}(2017)\citenamefont
  {Whittaker}, \citenamefont {Dzurnak}, \citenamefont {Egorov}, \citenamefont
  {Buonaiuto}, \citenamefont {Walker}, \citenamefont {Cancellieri},
  \citenamefont {Whittaker}, \citenamefont {Clarke}, \citenamefont {Gavrilov},
  \citenamefont {Skolnick},\ and\ \citenamefont
  {Krizhanovskii}}]{whittaker_polariton_2017}%
  \BibitemOpen
  \bibfield  {author} {\bibinfo {author} {C. Whittaker}, \bibinfo {author} {B.
  Dzurnak}, \bibinfo {author} {O. Egorov}, \bibinfo {author} {G. Buonaiuto},
  \bibinfo {author} {P. Walker}, \bibinfo {author} {E. Cancellieri}, \bibinfo
  {author} {D. Whittaker}, \bibinfo {author} {E. Clarke}, \bibinfo {author} {S.
  Gavrilov}, \bibinfo {author} {M. Skolnick}, \bibinfo {author} {D.
  Krizhanovskii},\ }\bibfield  {title} {\emph {\bibinfo {title} {Polariton
  {Pattern} {Formation} and {Photon} {Statistics} of the {Associated}
  {Emission}},\ }}\href {\doibase 10.1103/PhysRevX.7.031033} {\bibfield
  {journal} {\bibinfo  {journal} {Phys. Rev. X}\ }\textbf {\bibinfo {volume}
  {7}},\ \bibinfo {pages} {031033} (\bibinfo {year} {2017})}\BibitemShut
  {NoStop}%
\bibitem [{\citenamefont {Dominici}\ \emph
  {et~al.}(2015{\natexlab{a}})\citenamefont {Dominici}, \citenamefont {Petrov},
  \citenamefont {Matuszewski}, \citenamefont {Ballarini}, \citenamefont
  {Giorgi}, \citenamefont {Colas}, \citenamefont {Cancellieri}, \citenamefont
  {Fern\'{a}ndez}, \citenamefont {Bramati}, \citenamefont {Gigli},
  \citenamefont {Kavokin}, \citenamefont {Laussy},\ and\ \citenamefont
  {Sanvitto}}]{dominici_real-space_2015}%
  \BibitemOpen
  \bibfield  {author} {\bibinfo {author} {L. Dominici}, \bibinfo {author} {M.
  Petrov}, \bibinfo {author} {M. Matuszewski}, \bibinfo {author} {D.
  Ballarini}, \bibinfo {author} {M.~D. Giorgi}, \bibinfo {author} {D. Colas},
  \bibinfo {author} {E. Cancellieri}, \bibinfo {author} {B.~S. Fern\'{a}ndez},
  \bibinfo {author} {A. Bramati}, \bibinfo {author} {G. Gigli}, \bibinfo
  {author} {A. Kavokin}, \bibinfo {author} {F. Laussy}, \bibinfo {author} {D.
  Sanvitto},\ }\bibfield  {title} {\emph {\bibinfo {title} {Real-space collapse
  of a polariton condensate},\ }}\href {\doibase 10.1038/ncomms9993} {\bibfield
   {journal} {\bibinfo  {journal} {Nat. Commun.}\ }\textbf {\bibinfo {volume}
  {6}},\ \bibinfo {pages} {8993} (\bibinfo {year}
  {2015}{\natexlab{a}})}\BibitemShut {NoStop}%
\bibitem [{\citenamefont {Manni}\ \emph {et~al.}(2011)\citenamefont {Manni},
  \citenamefont {Lagoudakis}, \citenamefont {Liew}, \citenamefont {Andr\'{e}},\
  and\ \citenamefont {Deveaud-Pl\'{e}dran}}]{manni_spontaneous_2011}%
  \BibitemOpen
  \bibfield  {author} {\bibinfo {author} {F. Manni}, \bibinfo {author} {K.~G.
  Lagoudakis}, \bibinfo {author} {T.~C.~H. Liew}, \bibinfo {author} {R.
  Andr\'{e}}, \bibinfo {author} {B. Deveaud-Pl\'{e}dran},\ }\bibfield  {title}
  {\emph {\bibinfo {title} {Spontaneous {Pattern} {Formation} in a {Polariton}
  {Condensate}},\ }}\href {\doibase 10.1103/PhysRevLett.107.106401} {\bibfield
  {journal} {\bibinfo  {journal} {Phys. Rev. Lett.}\ }\textbf {\bibinfo
  {volume} {107}},\ \bibinfo {pages} {106401} (\bibinfo {year}
  {2011})}\BibitemShut {NoStop}%
\bibitem [{\citenamefont {Wertz}\ \emph {et~al.}(2010)\citenamefont {Wertz},
  \citenamefont {Ferrier}, \citenamefont {Solnyshkov}, \citenamefont {Johne},
  \citenamefont {Sanvitto}, \citenamefont {Lema\^{i}tre}, \citenamefont
  {Sagnes}, \citenamefont {Grousson}, \citenamefont {Kavokin}, \citenamefont
  {Senellart}, \citenamefont {Malpuech},\ and\ \citenamefont
  {Bloch}}]{wertz_spontaneous_2010}%
  \BibitemOpen
  \bibfield  {author} {\bibinfo {author} {E. Wertz}, \bibinfo {author} {L.
  Ferrier}, \bibinfo {author} {D.~D. Solnyshkov}, \bibinfo {author} {R. Johne},
  \bibinfo {author} {D. Sanvitto}, \bibinfo {author} {A. Lema\^{i}tre},
  \bibinfo {author} {I. Sagnes}, \bibinfo {author} {R. Grousson}, \bibinfo
  {author} {A.~V. Kavokin}, \bibinfo {author} {P. Senellart}, \bibinfo {author}
  {G. Malpuech}, \bibinfo {author} {J. Bloch},\ }\bibfield  {title} {\emph
  {\bibinfo {title} {Spontaneous formation and optical manipulation of extended
  polariton condensates},\ }}\href {\doibase 10.1038/nphys1750} {\bibfield
  {journal} {\bibinfo  {journal} {Nat. Phys.}\ }\textbf {\bibinfo {volume}
  {6}},\ \bibinfo {pages} {860} (\bibinfo {year} {2010})}\BibitemShut {NoStop}%
\bibitem [{\citenamefont {Ostrovskaya}\ \emph {et~al.}(2012)\citenamefont
  {Ostrovskaya}, \citenamefont {Abdullaev}, \citenamefont {Desyatnikov},
  \citenamefont {Fraser},\ and\ \citenamefont
  {Kivshar}}]{ostrovskaya_dissipative_2012}%
  \BibitemOpen
  \bibfield  {author} {\bibinfo {author} {E.~A. Ostrovskaya}, \bibinfo {author}
  {J. Abdullaev}, \bibinfo {author} {A.~S. Desyatnikov}, \bibinfo {author}
  {M.~D. Fraser}, \bibinfo {author} {Y.~S. Kivshar},\ }\bibfield  {title}
  {\emph {\bibinfo {title} {Dissipative solitons and vortices in polariton
  {Bose}-{Einstein} condensates},\ }}\href {\doibase
  10.1103/PhysRevA.86.013636} {\bibfield  {journal} {\bibinfo  {journal} {Phys.
  Rev. A}\ }\textbf {\bibinfo {volume} {86}},\ \bibinfo {pages} {013636}
  (\bibinfo {year} {2012})}\BibitemShut {NoStop}%
\bibitem [{\citenamefont {Sich}\ \emph {et~al.}(2012)\citenamefont {Sich},
  \citenamefont {Krizhanovskii}, \citenamefont {Skolnick}, \citenamefont
  {Gorbach}, \citenamefont {Hartley}, \citenamefont {Skryabin}, \citenamefont
  {Cerda-M\'{e}ndez}, \citenamefont {Biermann}, \citenamefont {Hey},\ and\
  \citenamefont {Santos}}]{sich_observation_2012}%
  \BibitemOpen
  \bibfield  {author} {\bibinfo {author} {M. Sich}, \bibinfo {author} {D.~N.
  Krizhanovskii}, \bibinfo {author} {M.~S. Skolnick}, \bibinfo {author} {A.~V.
  Gorbach}, \bibinfo {author} {R. Hartley}, \bibinfo {author} {D.~V. Skryabin},
  \bibinfo {author} {E.~A. Cerda-M\'{e}ndez}, \bibinfo {author} {K. Biermann},
  \bibinfo {author} {R. Hey}, \bibinfo {author} {P.~V. Santos},\ }\bibfield
  {title} {\emph {\bibinfo {title} {Observation of bright polariton solitons in
  a semiconductor microcavity},\ }}\href {\doibase 10.1038/nphoton.2011.267}
  {\bibfield  {journal} {\bibinfo  {journal} {Nat. Photon.}\ }\textbf {\bibinfo
  {volume} {6}},\ \bibinfo {pages} {50} (\bibinfo {year} {2012})}\BibitemShut
  {NoStop}%
\bibitem [{\citenamefont {Egorov}\ \emph {et~al.}(2010)\citenamefont {Egorov},
  \citenamefont {Gorbach}, \citenamefont {Lederer},\ and\ \citenamefont
  {Skryabin}}]{egorov_two-dimensional_2010}%
  \BibitemOpen
  \bibfield  {author} {\bibinfo {author} {O.~A. Egorov}, \bibinfo {author}
  {A.~V. Gorbach}, \bibinfo {author} {F. Lederer}, \bibinfo {author} {D.~V.
  Skryabin},\ }\bibfield  {title} {\emph {\bibinfo {title} {Two-{Dimensional}
  {Localization} of {Exciton} {Polaritons} in {Microcavities}},\ }}\href
  {\doibase 10.1103/PhysRevLett.105.073903} {\bibfield  {journal} {\bibinfo
  {journal} {Phys. Rev. Lett.}\ }\textbf {\bibinfo {volume} {105}},\ \bibinfo
  {pages} {073903} (\bibinfo {year} {2010})}\BibitemShut {NoStop}%
\bibitem [{\citenamefont {Ballarini}\ \emph {et~al.}(2013)\citenamefont
  {Ballarini}, \citenamefont {De~Giorgi}, \citenamefont {Cancellieri},
  \citenamefont {Houdr\'{e}}, \citenamefont {Giacobino}, \citenamefont
  {Cingolani}, \citenamefont {Bramati}, \citenamefont {Gigli},\ and\
  \citenamefont {Sanvitto}}]{ballarini_all-optical_2013}%
  \BibitemOpen
  \bibfield  {author} {\bibinfo {author} {D. Ballarini}, \bibinfo {author} {M.
  De~Giorgi}, \bibinfo {author} {E. Cancellieri}, \bibinfo {author} {R.
  Houdr\'{e}}, \bibinfo {author} {E. Giacobino}, \bibinfo {author} {R.
  Cingolani}, \bibinfo {author} {A. Bramati}, \bibinfo {author} {G. Gigli},
  \bibinfo {author} {D. Sanvitto},\ }\bibfield  {title} {\emph {\bibinfo
  {title} {All-optical polariton transistor},\ }}\href {\doibase
  10.1038/ncomms2734} {\bibfield  {journal} {\bibinfo  {journal} {Nat.
  Commun.}\ }\textbf {\bibinfo {volume} {4}},\ \bibinfo {pages} {1778}
  (\bibinfo {year} {2013})}\BibitemShut {NoStop}%
\bibitem [{\citenamefont {Sun}\ \emph {et~al.}(2015)\citenamefont {Sun},
  \citenamefont {Wade}, \citenamefont {Lee}, \citenamefont {Orcutt},
  \citenamefont {Alloatti}, \citenamefont {Georgas}, \citenamefont {Waterman},
  \citenamefont {Shainline}, \citenamefont {Avizienis}, \citenamefont {Lin},
  \citenamefont {Moss}, \citenamefont {Kumar}, \citenamefont {Pavanello},
  \citenamefont {Atabaki}, \citenamefont {Cook}, \citenamefont {Ou},
  \citenamefont {Leu}, \citenamefont {Chen}, \citenamefont {AsanoviÄ‡},
  \citenamefont {Ram}, \citenamefont {PopoviÄ‡},\ and\ \citenamefont
  {StojanoviÄ‡}}]{sun_single-chip_2015}%
  \BibitemOpen
  \bibfield  {author} {\bibinfo {author} {C. Sun} \emph {et~al.},\ }\bibfield
  {title} {\emph {\bibinfo {title} {Single-chip microprocessor that
  communicates directly using light},\ }}\href {\doibase 10.1038/nature16454}
  {\bibfield  {journal} {\bibinfo  {journal} {Nature}\ }\textbf {\bibinfo
  {volume} {528}},\ \bibinfo {pages} {534} (\bibinfo {year}
  {2015})}\BibitemShut {NoStop}%
\bibitem [{\citenamefont {Colas}\ \emph {et~al.}(2015)\citenamefont {Colas},
  \citenamefont {Dominici}, \citenamefont {Donati}, \citenamefont {Pervishko},
  \citenamefont {Liew}, \citenamefont {Shelykh}, \citenamefont {Ballarini},
  \citenamefont {de~Giorgi}, \citenamefont {Bramati}, \citenamefont {Gigli},
  \citenamefont {Valle}, \citenamefont {Laussy}, \citenamefont {Kavokin},\ and\
  \citenamefont {Sanvitto}}]{colas_polarization_2015}%
  \BibitemOpen
  \bibfield  {author} {\bibinfo {author} {D. Colas}, \bibinfo {author} {L.
  Dominici}, \bibinfo {author} {S. Donati}, \bibinfo {author} {A.~A.
  Pervishko}, \bibinfo {author} {T.~C. Liew}, \bibinfo {author} {I.~A.
  Shelykh}, \bibinfo {author} {D. Ballarini}, \bibinfo {author} {M. de~Giorgi},
  \bibinfo {author} {A. Bramati}, \bibinfo {author} {G. Gigli}, \bibinfo
  {author} {E.~d. Valle}, \bibinfo {author} {F.~P. Laussy}, \bibinfo {author}
  {A.~V. Kavokin}, \bibinfo {author} {D. Sanvitto},\ }\bibfield  {title} {\emph
  {\bibinfo {title} {Polarization shaping of {Poincar\'{e}} beams by polariton
  oscillations},\ }}\href {\doibase 10.1038/lsa.2015.123} {\bibfield  {journal}
  {\bibinfo  {journal} {Light Sci. Appl.}\ }\textbf {\bibinfo {volume} {4}},\
  \bibinfo {pages} {e350} (\bibinfo {year} {2015})}\BibitemShut {NoStop}%
\bibitem [{\citenamefont {Dominici}\ \emph {et~al.}(2014)\citenamefont
  {Dominici}, \citenamefont {Colas}, \citenamefont {Donati}, \citenamefont
  {Restrepo~Cuartas}, \citenamefont {De~Giorgi}, \citenamefont {Ballarini},
  \citenamefont {Guirales}, \citenamefont {L\'{o}pez Carre\~{n}o},
  \citenamefont {Bramati}, \citenamefont {Gigli}, \citenamefont {del Valle},
  \citenamefont {Laussy},\ and\ \citenamefont
  {Sanvitto}}]{dominici_ultrafast_2014}%
  \BibitemOpen
  \bibfield  {author} {\bibinfo {author} {L. Dominici}, \bibinfo {author} {D.
  Colas}, \bibinfo {author} {S. Donati}, \bibinfo {author} {J.~P.
  Restrepo~Cuartas}, \bibinfo {author} {M. De~Giorgi}, \bibinfo {author} {D.
  Ballarini}, \bibinfo {author} {G. Guirales}, \bibinfo {author} {J.~C.
  L\'{o}pez Carre\~{n}o}, \bibinfo {author} {A. Bramati}, \bibinfo {author} {G.
  Gigli}, \bibinfo {author} {E. del Valle}, \bibinfo {author} {F.~P. Laussy},
  \bibinfo {author} {D. Sanvitto},\ }\bibfield  {title} {\emph {\bibinfo
  {title} {Ultrafast {Control} and {Rabi} {Oscillations} of {Polaritons}},\
  }}\href {\doibase 10.1103/PhysRevLett.113.226401} {\bibfield  {journal}
  {\bibinfo  {journal} {Phys. Rev. Lett.}\ }\textbf {\bibinfo {volume} {113}},\
  \bibinfo {pages} {226401} (\bibinfo {year} {2014})}\BibitemShut {NoStop}%
\bibitem [{\citenamefont {Voronych}\ \emph {et~al.}(2017)\citenamefont
  {Voronych}, \citenamefont {Buraczewski}, \citenamefont {Matuszewski},\ and\
  \citenamefont {Stobi\'{n}ska}}]{voronych_numerical_2017}%
  \BibitemOpen
  \bibfield  {author} {\bibinfo {author} {O. Voronych}, \bibinfo {author} {A.
  Buraczewski}, \bibinfo {author} {M. Matuszewski}, \bibinfo {author} {M.
  Stobi\'{n}ska},\ }\bibfield  {title} {\emph {\bibinfo {title} {Numerical
  modeling of exciton-polariton {Bose}-{Einstein} condensate in a
  microcavity},\ }}\href {\doibase 10.1016/j.cpc.2017.02.021} {\bibfield
  {journal} {\bibinfo  {journal} {Comput. Phys. Commun.}\ }\textbf {\bibinfo
  {volume} {215}},\ \bibinfo {pages} {246} (\bibinfo {year}
  {2017})}\BibitemShut {NoStop}%
\bibitem [{\citenamefont {Bonaretti}\ \emph {et~al.}(2009)\citenamefont
  {Bonaretti}, \citenamefont {Faccio}, \citenamefont {Clerici}, \citenamefont
  {Biegert},\ and\ \citenamefont {Di~Trapani}}]{bonaretti_spatiotemporal_2009}%
  \BibitemOpen
  \bibfield  {author} {\bibinfo {author} {F. Bonaretti}, \bibinfo {author} {D.
  Faccio}, \bibinfo {author} {M. Clerici}, \bibinfo {author} {J. Biegert},
  \bibinfo {author} {P. Di~Trapani},\ }\bibfield  {title} {\emph {\bibinfo
  {title} {Spatiotemporal {Amplitude} and {Phase} {Retrieval} of {Bessel}-{X}
  pulses using a {Hartmann}-{Shack} {Sensor}},\ }}\href {\doibase
  10.1364/OE.17.009804} {\bibfield  {journal} {\bibinfo  {journal} {Opt.
  Express}\ }\textbf {\bibinfo {volume} {17}},\ \bibinfo {pages} {9804}
  (\bibinfo {year} {2009})}\BibitemShut {NoStop}%
\bibitem [{\citenamefont {Mugnai}\ \emph {et~al.}(2000)\citenamefont {Mugnai},
  \citenamefont {Ranfagni},\ and\ \citenamefont
  {Ruggeri}}]{mugnai_observation_2000}%
  \BibitemOpen
  \bibfield  {author} {\bibinfo {author} {D. Mugnai}, \bibinfo {author} {A.
  Ranfagni}, \bibinfo {author} {R. Ruggeri},\ }\bibfield  {title} {\emph
  {\bibinfo {title} {Observation of {Superluminal} {Behaviors} in {Wave}
  {Propagation}},\ }}\href {\doibase 10.1103/PhysRevLett.84.4830} {\bibfield
  {journal} {\bibinfo  {journal} {Phys. Rev. Lett.}\ }\textbf {\bibinfo
  {volume} {84}},\ \bibinfo {pages} {4830} (\bibinfo {year}
  {2000})}\BibitemShut {NoStop}%
\bibitem [{\citenamefont {Bowlan}\ \emph {et~al.}(2009)\citenamefont {Bowlan},
  \citenamefont {Valtna-Lukner}, \citenamefont {L\^{o}hmus}, \citenamefont
  {Piksarv}, \citenamefont {Saari},\ and\ \citenamefont
  {Trebino}}]{bowlan_measuring_2009}%
  \BibitemOpen
  \bibfield  {author} {\bibinfo {author} {P. Bowlan}, \bibinfo {author} {H.
  Valtna-Lukner}, \bibinfo {author} {M. L\^{o}hmus}, \bibinfo {author} {P.
  Piksarv}, \bibinfo {author} {P. Saari}, \bibinfo {author} {R. Trebino},\
  }\bibfield  {title} {\emph {\bibinfo {title} {Measuring the spatiotemporal
  field of ultrashort {Bessel}-{X} pulses},\ }}\href {\doibase
  10.1364/OL.34.002276} {\bibfield  {journal} {\bibinfo  {journal} {Opt.
  Lett.}\ }\textbf {\bibinfo {volume} {34}},\ \bibinfo {pages} {2276} (\bibinfo
  {year} {2009})}\BibitemShut {NoStop}%
\bibitem [{\citenamefont {Valtna-Lukner}\ \emph {et~al.}(2009)\citenamefont
  {Valtna-Lukner}, \citenamefont {Bowlan}, \citenamefont {L\^{o}hmus},
  \citenamefont {Piksarv}, \citenamefont {Trebino},\ and\ \citenamefont
  {Saari}}]{valtna-lukner_direct_2009}%
  \BibitemOpen
  \bibfield  {author} {\bibinfo {author} {H. Valtna-Lukner}, \bibinfo {author}
  {P. Bowlan}, \bibinfo {author} {M. L\^{o}hmus}, \bibinfo {author} {P.
  Piksarv}, \bibinfo {author} {R. Trebino}, \bibinfo {author} {P. Saari},\
  }\bibfield  {title} {\emph {\bibinfo {title} {Direct spatiotemporal
  measurements of accelerating ultrashort {Bessel}-type light bullets},\
  }}\href {\doibase 10.1364/OE.17.014948} {\bibfield  {journal} {\bibinfo
  {journal} {Opt. Express}\ }\textbf {\bibinfo {volume} {17}},\ \bibinfo
  {pages} {14948} (\bibinfo {year} {2009})}\BibitemShut {NoStop}%
\bibitem [{\citenamefont {Dominici}\ \emph
  {et~al.}(2015{\natexlab{b}})\citenamefont {Dominici}, \citenamefont
  {Dagvadorj}, \citenamefont {Fellows}, \citenamefont {Ballarini},
  \citenamefont {De~Giorgi}, \citenamefont {Marchetti}, \citenamefont
  {Piccirillo}, \citenamefont {Marrucci}, \citenamefont {Bramati},
  \citenamefont {Gigli}, \citenamefont {Szyma\'{n}ska},\ and\ \citenamefont
  {Sanvitto}}]{dominici_vortex_2015}%
  \BibitemOpen
  \bibfield  {author} {\bibinfo {author} {L. Dominici}, \bibinfo {author} {G.
  Dagvadorj}, \bibinfo {author} {J.~M. Fellows}, \bibinfo {author} {D.
  Ballarini}, \bibinfo {author} {M. De~Giorgi}, \bibinfo {author} {F.~M.
  Marchetti}, \bibinfo {author} {B. Piccirillo}, \bibinfo {author} {L.
  Marrucci}, \bibinfo {author} {A. Bramati}, \bibinfo {author} {G. Gigli},
  \bibinfo {author} {M.~H. Szyma\'{n}ska}, \bibinfo {author} {D. Sanvitto},\
  }\bibfield  {title} {\emph {\bibinfo {title} {Vortex and half-vortex dynamics
  in a nonlinear spinor quantum fluid},\ }}\href {\doibase
  10.1126/sciadv.1500807} {\bibfield  {journal} {\bibinfo  {journal} {Sci.
  Adv.}\ }\textbf {\bibinfo {volume} {1}},\ \bibinfo {pages} {e1500807}
  (\bibinfo {year} {2015}{\natexlab{b}})}\BibitemShut {NoStop}%
\bibitem [{\citenamefont {Rodrigues}\ \emph {et~al.}(2014)\citenamefont
  {Rodrigues}, \citenamefont {Kevrekidis}, \citenamefont
  {Carretero-Gonz\'{a}lez}, \citenamefont {Cuevas-Maraver}, \citenamefont
  {Frantzeskakis},\ and\ \citenamefont {Palmero}}]{rodrigues_nodeless_2014}%
  \BibitemOpen
  \bibfield  {author} {\bibinfo {author} {A.~S. Rodrigues}, \bibinfo {author}
  {P.~G. Kevrekidis}, \bibinfo {author} {R. Carretero-Gonz\'{a}lez}, \bibinfo
  {author} {J. Cuevas-Maraver}, \bibinfo {author} {D.~J. Frantzeskakis},
  \bibinfo {author} {F. Palmero},\ }\bibfield  {title} {\emph {\bibinfo {title}
  {From nodeless clouds and vortices to gray ring solitons and symmetry-broken
  states in two-dimensional polariton condensates},\ }}\href {\doibase
  10.1088/0953-8984/26/15/155801} {\bibfield  {journal} {\bibinfo  {journal}
  {J. Phys.: Condens. Matter}\ }\textbf {\bibinfo {volume} {26}},\ \bibinfo
  {pages} {155801} (\bibinfo {year} {2014})}\BibitemShut {NoStop}%
\bibitem [{\citenamefont {Lerario}\ \emph {et~al.}(2016)\citenamefont
  {Lerario}, \citenamefont {Ballarini}, \citenamefont {Fieramosca},
  \citenamefont {Cannavale}, \citenamefont {Genco}, \citenamefont {Mangione},
  \citenamefont {Gambino}, \citenamefont {Dominici}, \citenamefont {De~Giorgi},
  \citenamefont {Gigli},\ and\ \citenamefont
  {Sanvitto}}]{lerario_high-speed_2016}%
  \BibitemOpen
  \bibfield  {author} {\bibinfo {author} {G. Lerario}, \bibinfo {author} {D.
  Ballarini}, \bibinfo {author} {A. Fieramosca}, \bibinfo {author} {A.
  Cannavale}, \bibinfo {author} {A. Genco}, \bibinfo {author} {F. Mangione},
  \bibinfo {author} {S. Gambino}, \bibinfo {author} {L. Dominici}, \bibinfo
  {author} {M. De~Giorgi}, \bibinfo {author} {G. Gigli}, \bibinfo {author} {D.
  Sanvitto},\ }\bibfield  {title} {\emph {\bibinfo {title} {High-speed flow of
  interacting organic polaritons},\ }}\href {\doibase 10.1038/lsa.2016.212}
  {\bibfield  {journal} {\bibinfo  {journal} {Light Sci. Appl.}\ }\textbf
  {\bibinfo {volume} {6}},\ \bibinfo {pages} {e16212} (\bibinfo {year}
  {2016})}\BibitemShut {NoStop}%
\bibitem [{\citenamefont {Sinibaldi}\ \emph {et~al.}(2012)\citenamefont
  {Sinibaldi}, \citenamefont {Danz}, \citenamefont {Descrovi}, \citenamefont
  {Munzert}, \citenamefont {Schulz}, \citenamefont {Sonntag}, \citenamefont
  {Dominici},\ and\ \citenamefont {Michelotti}}]{sinibaldi_direct_2012}%
  \BibitemOpen
  \bibfield  {author} {\bibinfo {author} {A. Sinibaldi}, \bibinfo {author} {N.
  Danz}, \bibinfo {author} {E. Descrovi}, \bibinfo {author} {P. Munzert},
  \bibinfo {author} {U. Schulz}, \bibinfo {author} {F. Sonntag}, \bibinfo
  {author} {L. Dominici}, \bibinfo {author} {F. Michelotti},\ }\bibfield
  {title} {\emph {\bibinfo {title} {Direct comparison of the performance of
  {Bloch} surface wave and surface plasmon polariton sensors},\ }}\href
  {\doibase 10.1016/j.snb.2012.07.015} {\bibfield  {journal} {\bibinfo
  {journal} {Sens. Actuators B Chem.}\ }\textbf {\bibinfo {volume} {174}},\
  \bibinfo {pages} {292} (\bibinfo {year} {2012})}\BibitemShut {NoStop}%
\bibitem [{\citenamefont {Yu}\ \emph {et~al.}(2014)\citenamefont {Yu},
  \citenamefont {Barakat}, \citenamefont {Sfez}, \citenamefont {Hvozdara},
  \citenamefont {Di~Francesco},\ and\ \citenamefont
  {Peter~Herzig}}]{yu_manipulating_2014}%
  \BibitemOpen
  \bibfield  {author} {\bibinfo {author} {L. Yu}, \bibinfo {author} {E.
  Barakat}, \bibinfo {author} {T. Sfez}, \bibinfo {author} {L. Hvozdara},
  \bibinfo {author} {J. Di~Francesco}, \bibinfo {author} {H. Peter~Herzig},\
  }\bibfield  {title} {\emph {\bibinfo {title} {Manipulating {Bloch} surface
  waves in {2D}: a platform concept-based flat lens},\ }}\href {\doibase
  10.1038/lsa.2014.5} {\bibfield  {journal} {\bibinfo  {journal} {Light Sci.
  Appl.}\ }\textbf {\bibinfo {volume} {3}},\ \bibinfo {pages} {e124} (\bibinfo
  {year} {2014})}\BibitemShut {NoStop}%
\bibitem [{\citenamefont {Wang}\ \emph {et~al.}(2017)\citenamefont {Wang},
  \citenamefont {Wang}, \citenamefont {Zhang}, \citenamefont {Si},
  \citenamefont {Zhu}, \citenamefont {Du}, \citenamefont {Kou}, \citenamefont
  {Badugu}, \citenamefont {Rosenfeld}, \citenamefont {Lin}, \citenamefont
  {Wang}, \citenamefont {Ming}, \citenamefont {Yuan},\ and\ \citenamefont
  {Lakowicz}}]{wang_diffraction-free_2017}%
  \BibitemOpen
  \bibfield  {author} {\bibinfo {author} {R. Wang}, \bibinfo {author} {Y.
  Wang}, \bibinfo {author} {D. Zhang}, \bibinfo {author} {G. Si}, \bibinfo
  {author} {L. Zhu}, \bibinfo {author} {L. Du}, \bibinfo {author} {S. Kou},
  \bibinfo {author} {R. Badugu}, \bibinfo {author} {M. Rosenfeld}, \bibinfo
  {author} {J. Lin}, \bibinfo {author} {P. Wang}, \bibinfo {author} {H. Ming},
  \bibinfo {author} {X. Yuan}, \bibinfo {author} {J.~R. Lakowicz},\ }\bibfield
  {title} {\emph {\bibinfo {title} {Diffraction-{Free} {Bloch} {Surface}
  {Waves}},\ }}\href {\doibase 10.1021/acsnano.7b02358} {\bibfield  {journal}
  {\bibinfo  {journal} {ACS Nano}\ }\textbf {\bibinfo {volume} {11}},\ \bibinfo
  {pages} {5383} (\bibinfo {year} {2017})}\BibitemShut {NoStop}%
\bibitem [{\citenamefont {Lin}\ \emph {et~al.}(2012)\citenamefont {Lin},
  \citenamefont {Dellinger}, \citenamefont {Genevet}, \citenamefont {Cluzel},
  \citenamefont {de~Fornel},\ and\ \citenamefont
  {Capasso}}]{lin_cosine-gauss_2012}%
  \BibitemOpen
  \bibfield  {author} {\bibinfo {author} {J. Lin}, \bibinfo {author} {J.
  Dellinger}, \bibinfo {author} {P. Genevet}, \bibinfo {author} {B. Cluzel},
  \bibinfo {author} {F. de~Fornel}, \bibinfo {author} {F. Capasso},\ }\bibfield
   {title} {\emph {\bibinfo {title} {Cosine-{Gauss} {Plasmon} {Beam}: {A}
  {Localized} {Long}-{Range} {Nondiffracting} {Surface} {Wave}},\ }}\href
  {\doibase 10.1103/PhysRevLett.109.093904} {\bibfield  {journal} {\bibinfo
  {journal} {Phys. Rev. Lett.}\ }\textbf {\bibinfo {volume} {109}},\ \bibinfo
  {pages} {093904} (\bibinfo {year} {2012})}\BibitemShut {NoStop}%
\bibitem [{\citenamefont {Kishida}\ \emph {et~al.}(2000)\citenamefont
  {Kishida}, \citenamefont {Matsuzaki}, \citenamefont {Okamoto}, \citenamefont
  {Manabe}, \citenamefont {Yamashita}, \citenamefont {Taguchi},\ and\
  \citenamefont {Tokura}}]{kishida_gigantic_2000}%
  \BibitemOpen
  \bibfield  {author} {\bibinfo {author} {H. Kishida}, \bibinfo {author} {H.
  Matsuzaki}, \bibinfo {author} {H. Okamoto}, \bibinfo {author} {T. Manabe},
  \bibinfo {author} {M. Yamashita}, \bibinfo {author} {Y. Taguchi}, \bibinfo
  {author} {Y. Tokura},\ }\bibfield  {title} {\emph {\bibinfo {title} {Gigantic
  optical nonlinearity in one-dimensional {Mott}-{Hubbard} insulators},\
  }}\href {\doibase 10.1038/35016036} {\bibfield  {journal} {\bibinfo
  {journal} {Nature}\ }\textbf {\bibinfo {volume} {405}},\ \bibinfo {pages}
  {929} (\bibinfo {year} {2000})}\BibitemShut {NoStop}%
\bibitem [{\citenamefont {Deng}\ \emph {et~al.}(1999)\citenamefont {Deng},
  \citenamefont {Hagley}, \citenamefont {Wen}, \citenamefont {Trippenbach},
  \citenamefont {Band}, \citenamefont {Julienne}, \citenamefont {Simsarian},
  \citenamefont {Helmerson}, \citenamefont {Rolston},\ and\ \citenamefont
  {Phillips}}]{deng_four-wave_1999}%
  \BibitemOpen
  \bibfield  {author} {\bibinfo {author} {L. Deng}, \bibinfo {author} {E.~W.
  Hagley}, \bibinfo {author} {J. Wen}, \bibinfo {author} {M. Trippenbach},
  \bibinfo {author} {Y. Band}, \bibinfo {author} {P.~S. Julienne}, \bibinfo
  {author} {J.~E. Simsarian}, \bibinfo {author} {K. Helmerson}, \bibinfo
  {author} {S.~L. Rolston}, \bibinfo {author} {W.~D. Phillips},\ }\bibfield
  {title} {\emph {\bibinfo {title} {Four-wave mixing with matter waves},\
  }}\href {\doibase 10.1038/18395} {\bibfield  {journal} {\bibinfo  {journal}
  {Nature}\ }\textbf {\bibinfo {volume} {398}},\ \bibinfo {pages} {218}
  (\bibinfo {year} {1999})}\BibitemShut {NoStop}%
\bibitem [{\citenamefont {Gottesman}(1998)}]{gottesman_heisenberg_1998}%
  \BibitemOpen
  \bibfield  {author} {\bibinfo {author} {D. Gottesman},\ }\bibfield  {title}
  {\emph {\bibinfo {title} {The {Heisenberg} {Representation} of {Quantum}
  {Computers}},\ }}\href {http://arxiv.org/abs/quant-ph/9807006} {\bibfield
  {journal} {\bibinfo  {journal} {arXiv:quant-ph/9807006}\ } (\bibinfo {year}
  {1998})}\BibitemShut {NoStop}%
\end{thebibliography}

%

\end{document}